\documentstyle[prc,aps,epsf]{revtex}
\begin{document}
\draft

\title{ Three-body Halos. \\ IV. Momentum Distributions after Fragmentation }
\author{E.~Garrido, D.V.~Fedorov and A.S.~Jensen \\
Institute of Physics and Astronomy, \\
Aarhus University, DK-8000 Aarhus C, Denmark}
\date{\today}

\maketitle

\begin{abstract}

Momentum distributions of particles from nuclear break-up of fast
three-body halos are calculated incorporating effects of final state
interactions. The same two-body interactions between the particles are
used to calculate both the ground state structure and the final state
of the reaction processes. The ground state wave function reproduces
size and energy of the halo nucleus. First we give a general and
detailed description of the method. Then we investigate the effect of
final state interactions in neutron removal and core break-up reactions
for one- and two-dimensional momentum distributions.  We compute
specifically core and neutron momentum distributions from $^{11}$Li
fragmentation and compare those with available experimental data. We
conclude that $^{11}$Li must have a neutron-core relative state
containing a $p$-state admixture of 20\%-30\%. The fragmentation data
also strongly suggest that $^{10}$Li has an $s$-state at about 50 keV
and a $p$-state around 500 keV.

\vspace{5mm}


\end{abstract}

\vspace{5mm}

\section{Introduction} 
In the latest years a big effort has been made to investigate one of
the most striking new features in nuclear physics: halo nuclei.  They
are weakly bound and spatially extended systems where one or two particles 
have a high probability of being at distances
larger than the typical nuclear radius.  General properties and the
criteria for the ocurrence of nuclear halos have been discussed in
\cite{karsten,dima1,dima2,dima3} and a review of nuclear halos has
recently been published \cite{hansen}.

The appearence of the halo structure in light nuclei along the neutron
dripline \cite{detraz} gives special interest to this region of
nuclei.  The most general properties of them are well described by
few-body models, dividing the degrees of freedom into the
approximately frozen (core) and the active (halo) degrees of freedom.
In particular, special attention has been paid to Borromean systems
consisting of three-body bound systems, where all the two-body
subsystems are unbound \cite{hansen2,johannsen}. The prototypes
nuclear halos are $^6$He ($^4$He+n+n) and $^{11}$Li ($^9$Li+n+n), both
thoroughly discussed in a general theoretical framework in
\cite{zhukov}.

The ability to produce secondary beams of halo nuclei opens the
possibility of investigating their structure by measuring the momentum
distribution of ``particles" resulting from fragmentation reactions
\cite{kobayashi,anne,orr,orr2,zinser,nilsson,humbert}. However one
major problem in the interpretation of such measurements is the
inherent mixture of effects from the original structure of the
projectile and the reaction mechanism. Both should therefore be
properly incorporated in model calculations.

The simplest picture used to describe this kind of fragmentation
reactions is formulated in the sudden approximation, where one of the
three particles in the projectile is instantaneously removed by the
target, while the other two particles remain undisturbed. Clearly this
can only be justified for reaction times much shorter than the
characteristic time for the motion of the three particles in the
system.  Since the projectile is weakly bound this requirement is well
fulfilled for any high-energy beam. The observed momentum distribution
should then provide direct information about the three-body wave
function. The validity of this model as a first description of the
reaction has been discussed in the litterature, see for example
\cite{zhukov2,korshe,zhukov3}.

The connection between the wave function and the momentum
distributions is more complicated than originally anticipated when the
early experiments were designed.  It was found recently that momentum
distributions are highly affected by the final state interaction (FSI)
between the two particles remaining after the collision, especially
when low-lying resonances are present
\cite{zinser,korshe,barranco}. This essential ingredient must then be
incorporated into the model. The final state two-body interaction is
at the same time determining the three-body structure of the
projectile (halo nucleus), and therefore a consistent treatment of the
initial and final states is needed to allow reliable interpretations
of the experimental data.

This paper is number four in a series discussing the general
properties of three-body halo systems. In the first of these
\cite{dima2} we discussed Borromean systems, in the second
\cite{dima3} we extended the discussion to general three-body systems
and in the third \cite{dima4} we discussed the effects of finite spin
of the particles.  In this paper we investigate fragmentation
reactions of Borromean projectiles by use of the sudden
approximation. A brief report containing a few of the pertinent
results for $^{11}$Li is published \cite{gar96}. Our main purpose is
here to formulate a general procedure to incorporate the effects of
final state two-body interactions. This two-body interaction is
consistently included both in the final state wave function and in the
description of the three-body projectile. We apply the method in
detailed analyses of measured $^{11}$Li fragmentation reactions. We
thereby obtain essential information about the unbound $^{10}$Li
system.

The paper is organized as follows: The mathematical formalism is
developed in Section 2, which to facilitate the understanding is
divided in five subsections.  The general numerical results are
presented in Section 3, where we as illustration use parameters
relevant for $^{11}$Li fragmentation reactions. In Section 4 we
perform realistic calculations for $^{11}$ Li and compare to
experimental data. Finally, Section 5 contains a summary and the
conclusions. Some of the mathematical definitions are collected in an
appendix.

\section{Method}

To describe three-body halo fragmentation reactions we shall use the
sudden approximation. We are then assuming a process where a
high-energy three-body halo projectile instantaneously looses one of
the particles without disturbing the remaining two. We also assume a
light target and we shall therefore not consider Coulomb dissociation
process, which then only contribute marginally .

We work in the center of mass system of the three-body projectile and
denote by {\bf k} and {\bf q} the total and relative momentum of the
two remaining particles in the final state. The transition matrix of the
reaction in the sudden approximation is then given by
\begin{equation}
M(\mbox{\bf k}, \mbox{\bf q}) \propto \langle
 e^{i \mbox{\scriptsize \bf k} \cdot \mbox{\scriptsize \bf R} }
 e^{i \mbox{\scriptsize \bf q} \cdot \mbox{\scriptsize \bf r} }
           | \Psi \rangle  \; ,
\label{tran}
\end{equation}
where $\Psi$ is the three-body wave function, {\bf r} is the distance
between the two reamining particles, and {\bf R} is the distance between
the center of mass of the two-body system and the removed particle. 

  This transition matrix describes the final state as a plane wave,
which means that the interactions between the particles are
neglected. In principle all of them should be included in the
calculation, but under the experimental setup where one of the particles
suddenly is removed, only the interaction between the two
non-disturbed particles is present. To account for this final state
two-body interaction the plane wave $ e^{i \mbox{\scriptsize \bf q}
\cdot \mbox{\scriptsize \bf r} }$ has to be replaced by the
corresponding {\it distorted} two-body wave function $w(\mbox{\bf q},
\mbox{\bf r})$.

   In the following subsections we describe how we construct the
initial three-body halo wave function $\Psi$ as well as the distorted
two-body wave function $w(\mbox{\bf q}, \mbox{\bf r})$. Finally we
compute the transtion matrix and the momentum distributions.

\subsection{Initial Three-body Wave Function}

We shall use the Jacobi coordinates, basically defined as the relative
coordinates between two of the particles ({\bf x}) and between their
center of mass and the third particle ({\bf y}). Three sets of Jacobi
coordinates can be constructed and the precise definitions and the
corresponding sets of hyperspherical coordinates ($\rho$, $\alpha$,
$\Omega_x$, $\Omega_y$) are given in the appendix. One of the three
sets of hyperspherical coordinates is sufficient for a complete
description of the system, but the subtle correlations are easier to
describe by use of all three sets. The volume element is given by
$\rho^5 d\Omega d\rho$, where $d\Omega=\sin^2\alpha \cos^2\alpha
d\alpha d\Omega_x d\Omega_y$.

The total wave function $\Psi_{J M}$ of the three-body system (with
total spin $J$ and projection $M$) is written as a
sum of three components $\psi^{(i)}_{ J M}$, which in turn for each $\rho$
are expanded in a complete set of generalized angular functions
$\Phi^{(i)}_{n J M}(\rho,\Omega_i)$
\begin{equation}
\Psi_{J M}= \sum_{i=1}^3 \psi^{(i)}_{ J M}(\mbox{\bf x}_i, \mbox{\bf y}_i)
= \frac {1}{\rho^{5/2}} 
  \sum_n f_n(\rho)  
\sum_{i=1}^3 \Phi^{(i)}_{n J M}(\rho ,\Omega_i) \; ,
\label{tot}
\end{equation}
where the radial expansion coefficients $f_n(\rho)$ are component
independent and $\rho^{-5/2}$ is the phase space factor. 

These wave functions satisfy the three Faddeev equations \cite{bang,thomp}
\begin{equation}
(T-E)\psi^{(i)}_{ J M} + V_{jk} (\psi^{(i)}_{ J M}+\psi^{(j)}_{ J M}+
  \psi^{(k)}_{ J M})=0  \; ,
\label{fad}
\end{equation}
where $V_{jk}$ are the two-body interactions, $E$ is the total energy,
$T$ is the kinetic energy operator and $\{i,j,k\}$ is a cyclic
permutation of $\{1,2,3\}$.

The angular functions are now chosen for each $\rho$ as the
eigenfunctions of the angular part of the Faddeev equations:
\begin{equation} 
 {\hbar^2 \over 2m}\frac{1}{\rho^2}\hat\Lambda^2 \Phi^{(i)}_{n J M} 
 +V_{jk} (\Phi^{(i)}_{n J M}+\Phi^{(j)}_{n J M}+
                \Phi^{(k)}_{n J M})\equiv {\hbar^2 \over
2m}\frac{1}{\rho^2} \lambda_n(\rho) \Phi^{(i)}_{n J M}  \; ,
\label{ang}
\end{equation}
where $\{i,j,k\}$ again is a cyclic permutation of $\{1,2,3\}$, $m$ is
an arbitrary normalization mass, and $\hat\Lambda^2$ is the
$\rho$-independent part of the kinetic energy operator. The analytic
expressions for $\hat{\Lambda}^2$ and the kinetic energy operator can
for instance be found in \cite{dima3}.

The angular functions $\Phi^{(i)}_{n J M}(\rho ,\Omega_i)$ are
expanded in terms of the complete set of hyperspherical harmonics
$\mbox{\bf Y}_{\ell_x \ell_y}^{K L}(\alpha_i, \Omega_{x_i},
\Omega_{y_i})$, where the quantum number $K$ usually is called the
hypermoment, $\ell_x$ and $\ell_y$ are the orbital angular momenta
associated with {\bf x} and {\bf y}, and $L$ is the coupling of these
angular momenta. The result is
\begin{equation}
\Phi^{(i)}_{n J M}(\rho ,\Omega_i) = \sum_{K \ell_x \ell_y L s_x S}
C_{n K \ell_x \ell_y L s_x S}^{(i) }(\rho)
\left[ \mbox{\bf Y}_{\ell_x \ell_y}^{K L}(\alpha_i,\Omega_{x_i},\Omega_{y_i})
\otimes \chi_{s_x s_y S}^{(i)} \right]^{J M}  \; ,
\label{angcomp}
\end{equation}
where $C$ are expansion coefficients and $\chi_{s_x s_y S}^{(i)}$ is
the three-body spin function. The spins of the two particles,
connected by the {\bf x} coordinate, couple to the spin $s_x$, which
coupled to the spin $s_y$ of the third particle results in the total
spin $S$ of the three-body system.

The radial expansion coefficients $f_n(\rho)$ are obtained from 
a coupled set of ``radial'' differential equations \cite{dima3}, i.e.
\begin{equation} 
   \left(-\frac{\rm d ^2}{\rm d \rho^2}
   -{2mE\over\hbar^2}+ \frac{1}{\rho^2}\left( \lambda_n(\rho) +
  \frac{15}{4}\right) \right)f_n(\rho)
   +\sum_{n'}   \left(
   -2P_{n n'}{\rm d \over\rm d \rho}
   -Q_{n n'}
   \right)f_{n'}(\rho)
   =0  \; ,
\end{equation}
where the functions $P$ and $Q$ are defined as angular integrals:
\begin{equation}
   P_{n n'}(\rho)\equiv \sum_{i,j=1}^{3}
   \int d\Omega \Phi_n^{(i)\ast}(\rho,\Omega)
   {\partial\over\partial\rho}\Phi_{n'}^{(j)}(\rho,\Omega)  \; ,
\end{equation}
\begin{equation}
   Q_{n n'}(\rho)\equiv \sum_{i,j=1}^{3}
   \int d\Omega \Phi_n^{(i)\ast}(\rho,\Omega)
   {\partial^2\over\partial\rho^2}\Phi_{n'}^{(j)}(\rho,\Omega)  \; .
\end{equation}

Finally from eqs. (\ref{tot}) and (\ref{angcomp}) we obtain the
expression of the three-body wave function
\begin{equation}
\Psi_{J M} = \frac{1}{\rho^{5/2}} \sum_n f_n(\rho)
\sum_{i=1}^3 \sum_{K \ell_x \ell_y L s_x S}
C_{n K \ell_x \ell_y L s_x S}^{(i)}(\rho)
\left[ \mbox{\bf Y}_{\ell_x \ell_y}^{K L}(\alpha_i,\Omega_{x_i},\Omega_{y_i})
\otimes \chi_{s_x s_y S}^{(i) } \right]^{J M} \; ,
\label{tot3b}
\end{equation}
where the radial function
$f_n(\rho)$ and the coefficients $C_{n K \ell_x \ell_y L s_x
S}^{(i)}(\rho)$ must be computed numerically.

\subsection{Final-state Two-body Wave Function}
In the fragmentation reaction the target suddenly removes one of the
particles from the projectile and the remaining interaction in the
final state therefore acts between the two non-disturbed particles.
Assuming that the two-body interaction does not mix two-body states
with different spin $s_x$ and relative orbital angular momentum
$\ell_x$, we can expand the two-body wave function in partial waves,
see \cite{newton}, as 
\begin{eqnarray} 
 \label{2bexp} 
w^{s_x \sigma_x}(\mbox{\bf k}_x, \mbox{\bf x})= \sqrt{\frac{2}{\pi}}
\frac{1}{k_x x} \sum_{j_x \ell_x m_x} u_{\ell_x s_x}^{j_x} (k_x,x)
{\cal Y}_{j_x \ell_x s_x}^{m^*_x}(\Omega_x)
 \nonumber \\ 
\times \sum_{m_{\ell_x}=-\ell_x}^{\ell_x} 
\langle \ell_x m_{\ell_x} s_x \sigma_x|j_x m_x \rangle i^{\ell_x}
			      Y_{\ell_x m_{\ell_x}}(\Omega_{k_x})  \; ,
\end{eqnarray} 
where $\sigma_x$ is the spin projection of $s_x$, $u$ is
the radial and ${\cal Y}$ the angular distorted wave function. The
angles $\Omega_x$ and $\Omega_{k_x}$ define the direction of {\bf x}
and {\bf k}$_x$.  These assumptions are usually strictly valid. The
only exception arises from the tensor interaction and even then the
resulting mixing is often very small and therefore insignificant in the
present context.

The two-body space and momentum coordinates {\bf x} and {\bf k}$_x$
are defined consistently with the Jacobi coordinates introduced in the
three-body wave function (see appendix)
\begin{equation}
\mbox{\bf x} = \sqrt{\frac{\mu}{m}} \mbox{\bf r} \; ,
\hspace{1cm}
\mbox{\bf k}_x = \sqrt{\frac{m}{\mu}} \mbox{\bf p}_r  \; ,
\label{kp}
\end{equation}
where {\bf r} is the relative spatial coordinate and $\mbox{\bf p}_r$
the relative momentum. The mass $m$ is the arbitrary normalization
mass and $\mu$ is the reduced mass of the two-body system.

The radial functions $u_{\ell_x s_x}^{j_x}(k_x,x)$ are obtained numerically 
by solving the Schr\"odinger equation with the appropriate two-body potential
$\hat{V}(\mbox{\bf x})$
\begin{equation}
\frac{\partial^2}{\partial x^2} u_{\ell_x s_x}^{j_x}(k_x,x)+
\left( k_x^2 - \frac{2 m}{\hbar^2} V_{\ell_x s_x}^{j_x}(x) 
- \frac{\ell_x (\ell_x+1)}{x^2} \right)
u_{\ell_x s_x}^{j_x}(k_x,x)=0  \; ,
\label{schr}
\end{equation}
where
\begin{equation}
V_{\ell_x s_x}^{j_x}(x)=\int d\Omega_x 
                        {\cal Y}_{j_x \ell_x s_x}^{m_x^*}(\Omega_x)
  \hat{V}(\mbox{\bf x}) {\cal Y}_{j_x \ell_x s_x}^{m_x}(\Omega_x)  \; .
\end{equation}

When the interaction between the two particles is neglected the
solution of eq.(\ref{schr}) is $(k_x x) j_{\ell_x}(k_x x)$, and the
expansion in eq.(\ref{2bexp}) reduces to the usual expansion of a
plane wave in terms of spherical Bessel functions $j_{\ell}$.

\subsection{Transition Matrix}
To compute the transition matrix in eq.(\ref{tran}) it is convenient
to rewrite the three-body wave function in eq.(\ref{tot3b}) in terms
of the set of Jacobi coordinates where {\bf x} is related to the two
particles remaining after the fragmentation. This means that all three
Faddeev components in eq.(\ref{tot}) have to be expressed in terms of
the chosen set of Jacobi coordinates. The three-body wave function can
then be written
\begin{equation}
\Psi_{J M}(\mbox{\bf x}, \mbox{\bf y}) = \frac{1}{\rho^{5/2}} 
\sum_n f_n(\rho) \sum_{K \ell_x \ell_y L s_x S}
\tilde{C}_{n K \ell_x \ell_y L s_x S}(\rho)
\left[ \mbox{\bf Y}_{\ell_x \ell_y}^{K L}(\alpha, \Omega_x, \Omega_y)
\otimes \chi_{s_x s_y S} \right]^{J M}  \; ,
\label{totrot}
\end{equation}
where $\tilde{C}_{n K \ell_x \ell_y L s_x S}(\rho)$ are the
coefficients of the expansion after the transformation.

The calculation of the transition matrix involves the overlap of
eq.(\ref{totrot}) and the function $e^{i \mbox{\scriptsize \bf k}_y
\cdot \mbox{\scriptsize \bf y} } w^{s_x \sigma_x}(\mbox{\bf k}_x,
\mbox{\bf x})$ (see eqs.(\ref{tran}) and (\ref{2bexp}) and note that
$\mbox{\bf k} \cdot \mbox{\bf R} =
\mbox{\bf k}_y \cdot \mbox{\bf y}$, where $\mbox{\bf k}_y$ is the
momentum related to $\mbox{\bf y}$).  After integrating analytically
over $\Omega_x$ and $\Omega_y$ we get for each $\lambda_n$ the following
expression for the transition matrix:
\begin{eqnarray}
&& M^{J M}_{s_x \sigma_x s_y \sigma_y}(\mbox{\bf k}_x, \mbox{\bf k}_y) 
  \propto \frac{2}{\pi} \sum_{\ell_x m_{\ell_x} \ell_y m_{\ell_y}}
\sum_{j_x L S} I_{\ell_x s_x j_x}^{\ell_y L S}(\kappa, \alpha_\kappa)
Y_{\ell_x m_{\ell_x}}(\Omega_{k_x}) Y_{\ell_y m_{\ell_y}}(\Omega_{k_y})
    \nonumber \\ & &
 \times \sum_{m_x j_y m_y} (-1)^{J+2S-2M+\ell_y+s_y-s_x-\ell_x}
\hat{j}_x^2 \hat{j}_y^2 \hat{J} \hat{L} \hat{S}
\left(
    \begin{array}{ccc}
       J & j_x & j_y \\
       M & -m_x & -m_y
    \end{array}
    \right)
               \nonumber \\ & &
 \times \left(
    \begin{array}{ccc}
       j_y & \ell_y & s_y \\
       -m_y & m_{\ell_y} & \sigma_y
    \end{array}
    \right)
\left(
    \begin{array}{ccc}
       j_x & \ell_x & s_x \\
       -m_x & m_{\ell_x} & \sigma_x
    \end{array}
    \right)
\left\{
    \begin{array}{ccc}
       J & j_x & j_y \\
       L & \ell_x & \ell_y \\
       S & s_x  & s_y
    \end{array}
    \right\}  \; ,
\label{tm0}
\end{eqnarray}
where $\hat{a} \equiv \sqrt{2a+1}$, $\sigma_y$ is the
projection of $s_y$, $(\;)$ and $\{\;\}$ are the usual 3J and 9J
symbols, see \cite{brink}.  We have introduced the hyperspherical
coordinates in momentum space $\kappa=\sqrt{k_x^2+k_y^2}$,
$\alpha_\kappa=\arctan{(k_x/k_y)}$ and the function $I^{\ell_y L
S}_{\ell_x s_x j_x}(\kappa, \alpha_\kappa)$ is given by
\begin{eqnarray}
 && I^{\ell_y L S}_{\ell_x s_x j_x}(\kappa, \alpha_\kappa) = 
i^{\ell_x+\ell_y} \sum_K N_K^{\ell_x \ell_y}
\int_0^{\infty} \rho^{5/2} d\rho f(\rho) 
 \tilde{C}_{n K \ell_x \ell_y L s_x S}(\rho)
\label{ifun}
         \\ & \times &
\left[ \int_0^{\pi/2} d\alpha (\sin \alpha)^{\ell_x+2} (\cos \alpha)^{\ell_y+2}
P_\nu^{\ell_x+\frac{1}{2},\ell_y+\frac{1}{2}}\left(\cos(2\alpha)\right)
j_{\ell_y}(k_y y) \frac{1}{k_x x} u_{\ell_x s_x}^{j_x}(k_x, x) \right] 
   \nonumber \; ,
\end{eqnarray}
where $\nu=(K-\ell_x-\ell_y)/2$, $P_\nu^{\ell_x+\frac{1}{2},
\ell_y+\frac{1}{2}}$ is a Jacobi polynomial and
\begin{equation}
N_K^{\ell_x \ell_y}= \left[
\frac{\nu! (\nu+\ell_x+\ell_y+1)! 2 (K+2)}
        {\Gamma(\nu+\ell_x+\frac{3}{2}) \Gamma(\nu+\ell_y+\frac{3}{2})}
                \right]^{1/2}.
\end{equation}

When the interaction between the two particles in the final state is
ignored the integral over $\alpha$ in eq.(\ref{ifun}) can be carried out
analytically, i.e.
\begin{eqnarray}
\lefteqn{
\int _0^{\pi/2} d\alpha (\sin \alpha)^{\ell_x+2} (\cos \alpha)^{\ell_y+2}
P_\nu^{\ell_x+\frac{1}{2},\ell_y+\frac{1}{2}}\left(\cos(2\alpha)\right)
j_{\ell_y}(k_y y) j_{\ell_x}(k_x x)= } \nonumber\\ & & \times 
(-1)^\nu \frac{\pi}{2} (\sin\alpha_\kappa)^{\ell_x} 
(\cos\alpha_\kappa)^{\ell_y}
P_\nu^{\ell_x+\frac{1}{2} \ell_y+\frac{1}{2}}\left(\cos(2\alpha_\kappa)\right)
\frac{J_{K+2}(\kappa \rho)}{(\kappa \rho)^2} 
\end{eqnarray}

\subsection{Momentum Distributions}
The cross section or momentum distribution is now obtained by squaring
the transition matrix and subsequently averaging over initial states
and summing over final states:  
\begin{equation}
\frac{d^6\sigma}{d\mbox{\bf k}_x d\mbox{\bf k}_y} \propto
\sum_M \sum_{s_x \sigma_x \sigma_y}
|M^{J M}_{s_x \sigma_x s_y \sigma_y}(\mbox{\bf k}_x, \mbox{\bf k}_y)|^2 \; .
\label{mom}
\end{equation}

The volume element $d\mbox{\bf k}_x d\mbox{\bf k}_y$ is written 
\begin{equation}
d\mbox{\bf k}_x d\mbox{\bf k}_y=
k_x^\bot dk_x^\bot dk_x^\parallel d\varphi_{k_x} k_y^2 dk_y d\Omega_{k_y} \; ,
\end{equation}
where $k_x^\bot = k_x \sin{\theta_{k_x}}$, $k_x^\parallel=k_x
\cos{\theta_{k_x}}$, and $(\theta_{k_x}, \varphi_{k_x})$ are the polar
and azimuthal angles of the vector {\bf k}$_x$. The differential
cross section in eq.(\ref{mom}) should be integrated over all
unobserved variables.

Substituting eq.(\ref{tm0}) into (\ref{mom}) we obtain, after analytical
integration over $\Omega_{k_y}$ and $\varphi_{k_x}$, the
following expression for the three-dimensional differential cross
section or momentum distribution:
\begin{eqnarray}
\frac{d^3 \sigma}{dk_y dk_x^\bot dk_x^\parallel} & \propto &
k_y^2 k_x^\bot \frac{2}{\pi^2} 
\sum_{j_x \ell_x s_x }
\sum_{L S L^\prime S^\prime} \sum_{\ell_y j_y}
\hat{j_x}^2 \hat{j_y}^2 \hat{L} \hat{L^\prime} \hat{S} \hat{S^\prime}
I_{\ell_x s_x j_x}^{\ell_y L S}(\kappa, \alpha_\kappa)
I_{\ell_x s_x j_x}^{\ell_y L^\prime S^\prime}
(\kappa, \alpha_\kappa)
    \nonumber \\ & &
  \times \left\{
    \begin{array}{ccc}
       J & j_x & j_y \\
       L & \ell_x & \ell_y \\
       S & s_x & s_y
    \end{array}
    \right\}
\left\{
    \begin{array}{ccc}
       J & j_x & j_y \\
       L^\prime & \ell_x & \ell_y \\
       S^\prime & s_x & s_y
    \end{array}
    \right\}  \; .
\label{rel3}
\end{eqnarray}
By integrating numerically over $k_y$ and $k_x^\bot$, we get the
one-dimensional relative momentum $(k_x^\parallel)$ distribution of
the remaining particles.  By integrating over $k_y$ and
$k_x^\parallel$, we get instead the two-dimensional relative momentum
($k_x^\bot$) distribution.  

It should be noted that we have not specified any coordinate system
and the axes $\mbox{\bf x}$, $\mbox{\bf y}$, and $\mbox{\bf z}$ are
therefore completely arbitrary.  Thus, in the sudden approximation the
longitudinal and transverse momentum distributions are identical.

\subsection{Transformation to the Center of Mass of the Three-body
System}

Comparison with the measured momentum distributions requires
transformation to the center of mass system of the projectile. To do
this we construct the momentum $\mbox{\bf p}$ of one of the
particles in the final state relative to the center of mass of the
projectile as a linear combination of $\mbox{\bf k}_x$ and $\mbox{\bf
k}_y$
\begin{equation}
\mbox{\bf p} = a_i \mbox{\bf k}_x + b_i \mbox{\bf k}_y \; .
\label{trans}
\end{equation}
If the Jacobi coordinate {\bf x} refers to particles 1 and 2, then $a$
and $b$ take the set of values (see eqs.(\ref{pi}) and (\ref{pj}))
\begin{equation}
a_1=-\left( \frac{1}{m} \frac{m_1 m_2}{m_1+m_2} \right)^{1/2},
\hspace{1cm}
b_1=\frac{m_1}{m_1+m_2} \left(\frac{1}{m} 
             \frac{(m_1+m_2)m_3}{m_1+m_2+m_3} \right)^{1/2} \; ,
\end{equation}

\begin{equation}
a_2= \left( \frac{1}{m} \frac{m_1 m_2}{m_1+m_2} \right)^{1/2},
\hspace{1cm}
b_2=\frac{m_2}{m_1+m_2} \left(\frac{1}{m}
             \frac{(m_1+m_2)m_3}{m_1+m_2+m_3} \right)^{1/2} \; ,
\label{ab2}
\end{equation}
when we compute the momentum distributions of particle 1 and 2,
respectively.

Using the momentum $\mbox{\bf p}$ of one of the particles as the variable
instead of $\mbox{\bf k}_x$ we obtain 
the relation
\begin{equation}
\frac{d^6\sigma}{d\mbox{\bf p} d\mbox{\bf k}_y} = 
  \frac{1}{a_i^3} \frac{d^6\sigma}{d\mbox{\bf k}_x d\mbox{\bf k}_y}
  \propto \frac{1}{a_i^3}
\sum_M \sum_{s_x \sigma_x \sigma_y}
|M^{J M}_{s_x \sigma_x s_y \sigma_y}(\mbox{\bf k}_x, \mbox{\bf k}_y)|^2 \; ,
\label{rot}
\end{equation}
where $a_i^3$ arises from the Jacobi determinant for the
transformation.

As before we must integrate eq.(\ref{rot}) over the unobserved
quantities, i.e. $\mbox{\bf k}_y$ and some of the components of {\bf
p}.  It is then convenient to use both the initial (arbitrary)
coordinate system and a rotated system where the $z$-axis is along
{\bf p}. The volume element is then written as
\begin{equation}
d\mbox{\bf p} d\mbox{\bf k}_y=
p^\bot dp^\bot dp^\parallel d\varphi_p k_y^2 dk_y d\Omega_{k_y}^\prime \; ,
\end{equation}
where $p^\bot = p \sin{\theta_p}$, $p^\parallel=p \cos{\theta_p}$,
$(\theta_p, \varphi_p)$ define the direction of {\bf p} in the initial
coordinate system and the angles
$\Omega_{k_y}^\prime=(\theta_{k_y}^\prime, \varphi_{k_y}^\prime)$ give
the direction of $\mbox{\bf k}_y$ in the rotated system.

We express the transition matrix in eq.(\ref{tm0}) as a function of
{\bf p} referred to the initial system and $\mbox{\bf k}_y$ referred
to the rotated system. Then we need the relations
\begin{equation}
Y_{\ell m}(\theta_{k_y}, \varphi_{k_y})= \sum_{m^\prime} 
{\cal D}_{m m^\prime}^{\ell^*}(\varphi_p, \theta_p, 0) 
Y_{\ell m^\prime}(\theta_{k_y}^\prime, \varphi_{k_y}^\prime) \; ,
\end{equation}
\begin{equation}
Y_{\ell m}(\theta_{k_x}, \varphi_{k_x})= \sum_{m^\prime} 
{\cal D}_{m m^\prime}^{\ell^*}(\varphi_p, \theta_p, 0)
Y_{\ell m^\prime}(\theta_{k_x}^\prime, \varphi_{k_x}^\prime)  \; ,
\end{equation}
where ${\cal D}(\varphi_p, \theta_p, 0)$ is the rotation matrix (the
${\cal D}$-functions) defined in \cite{brink}, and the angles
($\theta_{k_x}^\prime, \varphi_{k_x}^\prime$) defining the direction
of $\mbox{\bf k}_x$ in the rotated system are given by
\begin{equation}
\cos{\theta_{k_x}^\prime} =
 \frac{p-b_i k_y \cos{\theta_{k_y}^\prime}}{a_i k_x}\; ,
\label{cosx}
\end{equation}
\begin{equation}
\varphi_{k_x}^\prime= \left\{
   \begin{array}{lcl}
           \varphi_{k_y}^\prime &\mbox{if}& a_i/b_i < 0 \\
           \varphi_{k_y}^\prime+\pi &\mbox{if}& a_i/b_i > 0
   \end{array}
               \right.  \; 
\end{equation}

Finally, we need to express $\kappa$ and $\alpha_{\kappa}$ in terms of
the new variables $p$, $k_y$ and $\theta_{k_y}^\prime$, i.e.
\begin{equation}
\kappa^2=\frac{1}{a_i^2} \left(
 p^2 + b_i^2 k_y^2 - 2 b_i p k_y \cos{\theta_{k_y}^\prime} \right) + k_y^2 \; ,
\label{kappa}
\end{equation}
\begin{equation}
\tan{\alpha_\kappa} = \frac{k_x}{k_y}=
\frac{(p^2 + b_i^2 k_y^2 - 2 b_i p k_y \cos{\theta_{k_y}^\prime})^{1/2}}
 {|a_i| k_y} \; , 
\hspace{1cm} 0 \leq \alpha_\kappa \leq \frac{\pi}{2} \; .
\label{akappa}
\end{equation}

The numerically obtained function $I^{\ell_y L S}_{\ell_x s_x
j_x}(\kappa, \alpha_\kappa)$ in eq.(\ref{tm0}) depends on
$\theta_{k_y}^\prime$ and only the integrations over $\varphi_p$ and
$\varphi_{k_y}^\prime$ in eq.(\ref{rot}) can be done analytically. We
then obtain the four-dimensional differential cross section or
momentum distribution relative to the center of mass of the projectile
\begin{small}
\begin{eqnarray}
&& \frac{d^4\sigma}{dk_y d\theta^\prime_{k_y} dp^\bot dp^\parallel}
    \propto   k_y^2 p^\bot \frac{1}{a_i^3 \pi^2} \sin{\theta^\prime_{k_y}}
\sum_{\ell_x s_x j_x L S}
\sum_{\ell_x^\prime j_x^\prime L^\prime S^\prime}
\sum_{\ell_y j_y \ell_y^\prime j_y^\prime}
(-1)^{s_y-s_x+j_y^\prime+j_y+J}
         \nonumber \\ & & \times
\hat{j_x}^2 \hat{j^\prime_x}^2 \hat{j_y}^2 \hat{j_y^\prime}^2 
\hat{\ell_x} \hat{\ell^\prime_x} \hat{\ell_y}
\hat{\ell^\prime_y} \hat{L} \hat{L^\prime} \hat{S} \hat{S^\prime}
I_{\ell_x s_x j_x}^{\ell_y L S}(\kappa, \alpha_\kappa)
I_{\ell_x^\prime s_x j_x^\prime}^{\ell_y^\prime L^\prime S^\prime}
(\kappa, \alpha_\kappa)
\left\{
    \begin{array}{ccc}
       J & j_x & j_y \\
       L & \ell_x & \ell_y \\
       S & s_x & s_y
    \end{array}
    \right\}
         \nonumber \\ & & \times
\left\{
    \begin{array}{ccc}
       J & j_x^\prime & j_y^\prime \\
       L^\prime & \ell_x^\prime & \ell_y^\prime \\
       S^\prime & s_x & s_y
    \end{array}
    \right\}
\sum_{L_x} \hat{L_x}^2 
\left(
    \begin{array}{ccc}
       \ell_x & \ell_x^\prime & L_x \\
       0 & 0 & 0
    \end{array}
    \right) 
\left(
    \begin{array}{ccc}
       \ell_y & \ell_y^\prime & L_x \\
       0 & 0 & 0
    \end{array}
    \right) 
\left\{
    \begin{array}{ccc}
       j_y & j^\prime_y & L_x \\
       \ell^\prime_y & \ell_y & s_y
    \end{array}
    \right\} 
         \nonumber \\ & & \times
\left\{
    \begin{array}{ccc}
       j_x & j^\prime_x & L_x \\
       \ell^\prime_x & \ell_x & s_x
    \end{array}
    \right\} 
\left\{
    \begin{array}{ccc}
       j_x & j^\prime_x & L_x \\
       j^\prime_y & j_y & J
    \end{array}
    \right\} 
\sum_{N_x} (-1)^{N_x}
P_{L_x}^{N_x}(\cos{\theta^\prime_{k_x}})
P_{L_x}^{-N_x}(\cos{\theta^\prime_{k_y}})  \; ,
\label{cm3}
\end{eqnarray}
\end{small}
where we assumed that $a_i<0$. When $a_i>0$, the summation over $N_x$
should include an extra factor $(-1)^{N_x}$.

Three integrations must be done numerically in order to get the
one-dimensional ($p^\parallel$) and two-dimensional $(p^\bot)$
momentum distributions of one of the particles in the final state
relative to the center of mass of the projectile.

If we assume that one of the particles in the final state, say
particle 1, has infinite mass, then the momentum distribution of
particle 2 relative to the three-body center of mass should coincide
with the momentum distribution relative to particle 1. In other words,
eq.(\ref{rel3}) should be recovered by integrating eq.(\ref{cm3}) over
$\theta^\prime_{k_y}$. Then $b_i=0$ (see eq.(\ref{ab2})), $\mbox{\bf p}
= a_i\mbox{\bf k}_x$ and therefore $\theta^\prime_{k_x}=0$ and
$P_{L_x}^{N_x}(1)=\delta_{N_x,0}$.  Now $\kappa$ and $\alpha_\kappa$
are independent of $\theta^\prime_{k_y}$ (eqs.(\ref{kappa}) and
(\ref{akappa})) and the integration over $\theta^\prime_{k_y}$ can easily
be done analytically leading directly to eq.(\ref{rel3}).

\section{Numerical results}

In this section we apply the method to the fragmentation of
two-neutron halo nuclei (core+n+n). Two different kinds of processes
will be considered: neutron removal and core break-up reactions.  In
the first case one of the neutrons is removed by the target, and the
neutron-core interaction in the final state must be considered. In the
second case, where the core is violently removed from the projectile,
the neutron-neutron interaction is involved in the final state.

We take parameters corresponding to $^{11}$Li in our examples.  We
first specify the two-body potentials used in the calculation. We then
investigate the momentum distributions and how they are affected by
final state interactions and the structure of the projectile. 

\subsection{Two-body Potentials}

Loosely bound systems are mainly sensitive to the low-energy properties
of the potentials. We shall therefore use relatively simple potentials
reproducing the available low-energy scattering data and still allowing
extensive three-body calculations.

For the neutron-neutron interaction in the singlet $s$-wave we use the
gaussian potential introduced in ref.\cite{johannsen}. The extension
to the triplet $p$-wave is made by including spin-spin, spin-orbit,
and tensor terms. The neutron-neutron potential has then the form
\begin{equation}
V_{nn}=\left( V_c+V_{ss} \mbox{\bf s}_{n1} \cdot \mbox{\bf s}_{n2}
+V_T \hat{S}_{12}+V_{so} \mbox{\bf l}_{nn} \cdot \mbox{\bf s}_{nn}
\right) \exp \left[ - (r/b_{nn})^2 \right] \; ,
\end{equation}
where $\mbox{\bf s}_{n1}$ and $\mbox{\bf s}_{n2}$ are the spins of the
two neutrons, $\mbox{\bf s}_{nn}=\mbox{\bf s}_{n1}+\mbox{\bf s}_{n2}$,
$\mbox{\bf l}_{nn}$ is the relative neutron-neutron orbital angular
momentum and $\hat{S}_{12}$ is the ususal tensor operator.

The strength parameters $V_c$, $V_{ss}$, $V_T$, and $V_{so}$, and the
range parameter $b_{nn}$ are adjusted to reproduce the following 
scattering lengths $a$ and the $s$-wave effective range $r_e$ \cite{dumbrajs}: 
\begin{eqnarray}
a(^1S_0)=18.8 \mbox{ fm } \; , & r_e(^1S_0)=2.76 \mbox{ fm }  \; ,  &  \\ 
a(^3P_0)=3.6 \mbox{ fm } \; ,  &
a(^3P_1)=-2.0 \mbox{ fm }  \; ,  &
a(^3P_2)=0.30 \mbox{ fm } \; .
\end{eqnarray}
We then obtain values for the parameters of the potential
\begin{eqnarray}
V_c=2.92 \mbox{ MeV } \; , & V_{ss}=45.22 \mbox{ MeV }  \; , \\  
V_T=26.85 \mbox{ MeV } \; , & V_{so}=-12.08 \mbox{ MeV }  \; , \\ 
b_{nn}=1.8 \mbox{ fm } \; . &       
\end{eqnarray}

For the neutron-core potential we also assume a gaussian shape and we use 
the parametrization
\begin{eqnarray}
V_{nc}^{(s)}&=&V_s (1+\gamma_s \mbox{\bf s}_c \cdot \mbox{\bf s}_n)
   \exp \left[ - (r/b_{nc})^2 \right], \label{pots} \\
V_{nc}^{(l)}&=&(V_l + V_{so}^{(l)} \mbox{\bf l}_{nc} \cdot \mbox{\bf s}_{nc})
   \exp \left[ - (r/b_{nc})^2 \right] \label{potl}  \; ,
\end{eqnarray}
where $\mbox{\bf s}_c$ and $\mbox{\bf s}_n$ are the spins of the core
and the neutron, $\mbox{\bf s}_{nc}=\mbox{\bf s}_c+\mbox{\bf s}_n$
and $\mbox{\bf l}_{nc}$ is the relative neutron-core orbital angular
momentum.  This choice is convenient when we adjust the all-decisive
energy positions of the virtual states and the resonances in the
neutron-core subsystem.  The spin splitting term $\mbox{\bf s}_c \cdot
\mbox{\bf s}_n$ is for simplicity only introduced in the $s$-state,
but it could as well be included for other partial waves.

\begin{figure}[t]
\epsfxsize=12cm
\epsfysize=7cm
\epsfbox[-100 400 450 750]{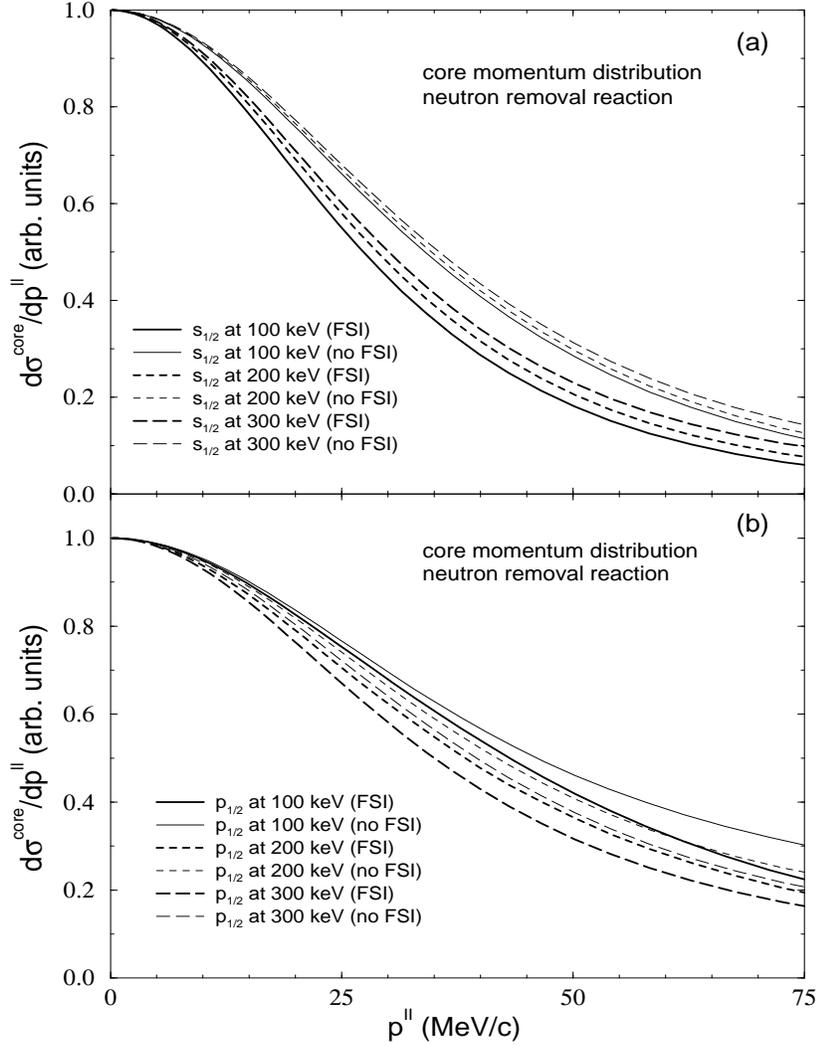}
\vspace{7.5cm}
\caption{\protect\small
One-dimensional core momentum distribution for a neutron
removal process in $^{11}$Li fragmentation. Thin curves correspond to
calculations without FSI, while FSI are included for the thick
curves. In the upper part the virtual $s_{1/2}$-state is placed at 100
keV (solid curve), 200 keV (short-dashed curve), and 300 keV
(long-dashed curve), respectively. In the lower part the
$p_{1/2}$-resonance is placed at 100 keV (solid curve), 200 keV
(short-dashed curve) and 300 keV (long-dashed curve),
respectively. The momentum of the core is referred to the center of
mass of the $^{11}$Li projectile. The $^{11}$Li spin is assumed to
be zero.
} 
\label{1}
\end{figure}

\begin{table}
\hspace*{0.8cm}
\begin{tabular}{cc|cccc|cc} 
$E_{s_{1/2}}$  & $E_{p_{1/2}}$  & $V_s$  & $V_{l=1}$  & $V_{so}$  & $b_{nc}$  &
       $s$-wave & $p$-wave  \\
  (keV) & (keV) & (MeV) & (MeV) & (MeV) & (fm) & (\%) & (\%) \\ \hline 
  100 & $>$5000 & $-7.80$ & $-4.5$ & 10.80 & 2.55 & 97 & 3 \\
  200 & 1700  & $-7.14$ & $-4.5$ & 30.87 & 2.55 & 84 & 16 \\
  300 &  860  & $-6.64$ & $-4.5$ & 33.60 & 2.55 & 73 & 27 \\
  810 &  300  & $-4.94$ & $-4.5$ & 36.42 & 2.55 & 49 & 51 \\
 1280 &  200  & $-3.85$ & $-4.5$ & 37.03 & 2.55 & 36 & 64 \\
 2600 &  100  & $-1.46$ & $-4.5$ & 37.67 & 2.55 & 22 & 78 \\
\end{tabular}
\vspace{0.3cm}
\caption{
Energy positions of the virtual $s_{1/2}$-state
and the $p_{1/2}$-resonance in the neutron-$^9$Li subsystem (first and
second column) and the parameters of the neutron-core potential for each 
case. The last two columns give the $s$- and $p$-wave content in the 
neutron-$^9$Li subsystem in the three-body wave function of $^{11}$Li. 
                 }
\end{table}

\subsection{Momentum Distributions for Spin Zero Core}
We analyze the general properties of momentum distributions and the
effects of final state interactions by taking $^{11}$Li as an
example. We include $s$- and $p$-waves in the Faddeev components and
neglect all higher orbital angular momenta, since their contributions
are expected to be exceedingly small.

We first consider the simpler case where the spin of the $^{9}$Li-core
is assumed to be zero. The spin quantum numbers of $^{10}$Li and $^{11}$Li
are correspondingly 1/2 and 0. The range parameter $b_{nc}$ in eqs.
(\ref{pots}) and (\ref{potl}) is assumed to be 2.55 fm
\cite{johannsen}.  The strength parameter $V_s$ is used to modify the
position of the virtual $s_{1/2}$-state in $^{10}$Li. The parameter
$V_{l=1}$ is chosen as $-4.5$ MeV and the spin-orbit parameter $V_{so}$
must be large to reproduce the experimental binding energy ($295 \pm
35$ keV \cite{young}) and root mean square radius ($3.1 \pm 0.3$ fm
\cite{tanihata}) of $^{11}$Li. With this choice of the parameters we
have a simple structure for the three-body system, where only the
positions of the virtual $s_{1/2}$-state and the $p_{1/2}$-resonance
in $^{10}$Li are relevant. In particular, then a $p_{3/2}$-resonance
appears at a very large energy. We can now investigate how the
momentum distributions are influenced by final state interactions for
different structures of the neutron-core subsystem.

Comparing the cases of low-lying virtual $s$-states and low-lying
$p$-resonances are of special interest. Six different situations are
considered: The virtual $s_{1/2}$-state in $^{10}$Li is at 100, 200,
and 300 keV, and the $p_{1/2}$-resonance is at 100, 200, and 300
keV. Because of the requirement of reproducing the experimental
binding energy and rms radius in $^{11}$Li both the virtual
$s_{1/2}$-state and the $p_{1/2}$-resonance can not simultaneously be
at a low energy \cite{dima4}. In table~I we show the values of the
parameters used in the neutron-core interaction as well as the
resulting $s$- and $p$-wave content in the neutron-core subsystem of
the total three-body wave function.

In fig.\ref{1} we show the one-dimensional core momentum distribution for a
neutron removal reaction measured in the center of mass of the
three-body projectile. To make the comparison easier all the curves
have been scaled to the same maximum. The thin curves are
the calculations without final state interactions while the thick
curves are obtained with final state interactions. In the upper part
of the figure the virtual $s_{1/2}$-state has been placed at 100 keV
(solid curve), 200 keV (short-dashed curve) and 300 keV (long-dashed
curve), and in the lower part the $p_{1/2}$-resonance energy is at
100 keV (solid curve), 200 keV (short-dashed curve) and 300 keV
(long-dashed curve). The full width at half maximum (FWHM) for the
momentum distributions in the figure are given in the second and third
columns of table II.

From fig.\ref{1} and table II we observe the two important features. First,
for a given neutron-$^9$Li structure final state interactions always
make the momentum distributions narrower. The change is not drastic,
due to the large mass of the core, but it is still important. The
largest effects appear for low-lying virtual $s_{1/2}$-states, where the
FWHM decreases by 15-19\%. For low-lying $p$-resonances the FWHM
decreases by less than 10\%. Thus, the larger the $s$-wave content in
the neutron-core subsystem, the larger the effect of final state
interactions.

Secondly, the width of the momentum distributions increases from 55
MeV/c, when the virtual $s_{1/2}$-state is at 100 keV, to 87 MeV/c,
when the $p_{1/2}$-resonance is at 100 keV. The corresponding numbers
without final state interactions are 68 MeV/c and 92 MeV/c. Comparison
with the experimental value in table II strongly suggests a low-lying
virtual $s$-state in $^{10}$Li.  The corresponding $s$-wave content
decreases from 97\% for the narrowest to only 22\% for the broadest
distribution.  Thus the larger the $s$-wave content in the initial
neutron-core subsystem of $^{11}$Li, the narrower the momentum
distribution.

\begin{table}
\hspace*{1.2cm}
\begin{tabular}{c|cc|ccc}
 Energy & \multicolumn{2}{c}{core} & \multicolumn{3}{c}{neutron} \\ \cline{2-6}
 (keV)  & no FSI & FSI$^{\mbox{\scriptsize (n-c)}}$  
        & no FSI & FSI$^{\mbox{\scriptsize (n-c)}}$ 
                 & FSI$^{\mbox{\scriptsize (n-n)}}$ \\
\cline{1-6}
$E_{s_{1/2}}=100$ &   68  & 55 &  51   & 28 & 39 \\
$E_{s_{1/2}}=200$ &   69  & 58 &  53   & 34 & 40 \\
$E_{s_{1/2}}=300$ &   71  & 60 &  55   & 40 & 41 \\
$E_{p_{1/2}}=300$ &   79  & 70 &  68   & 48 & 46 \\
$E_{p_{1/2}}=200$ &   84  & 77 &  77   & 43 & 48 \\
$E_{p_{1/2}}=100$ &   92  & 87 &  94   & 31 & 55 \\
Experimental  &         & $49\pm 3^{\mbox{\scriptsize (a)}}$ & 
            & 25-35$^{\mbox{\scriptsize (b)}}$
            & $43\pm 3^{\mbox{\scriptsize (c)}}$ \\
\end{tabular}
\vspace{0.3cm}
\caption{ Full width at half maximum in MeV/c of the one-dimensional
core and neutron momentum distributions from $^{11}$Li fragmentation
for different positions of the virtual $s_{1/2}$-state and the
$p_{1/2}$-resonance. Columns 2 and 3 refer to core momentum
distributions, while 4 to 6 refer to neutron momentum distributions.
The label (n-c) indicates that final state interactions between $^9$Li
and the neutron have been included in this neutron removal process. The
label (n-n) indicates that final state interactions between the two
neutrons have been included (core break-up) [a] Data from
\protect\cite{humbert}.  [b] See, for instance, \protect\cite{anne}.
[c] Data from \protect\cite{nilsson}.  }
\end{table}

These two facts can be explained by the effects of the centrifugal
barrier. A high $p$-wave content in the neutron-core subsystem
increases the importance of the centrifugal barrier, see
eq.(\ref{schr}). Consequently the effects of distortion from the
short-range potentials is smaller for $p$-waves than for $s$-waves.
Therefore the effects of final state interactions decreases when the
$p$-wave content increases. Furthermore, the probability for the
neutron to be outside the range of the two-body potential drastically
decreases for $p$-waves \cite{karsten,dima2}. This decreases the
spatial extension of the initial system when the $p$-wave content
increases, and the momentum distribution is then broader.

\begin{figure}[t]
\epsfxsize=12cm
\epsfysize=7cm
\epsfbox[-100 400 450 750]{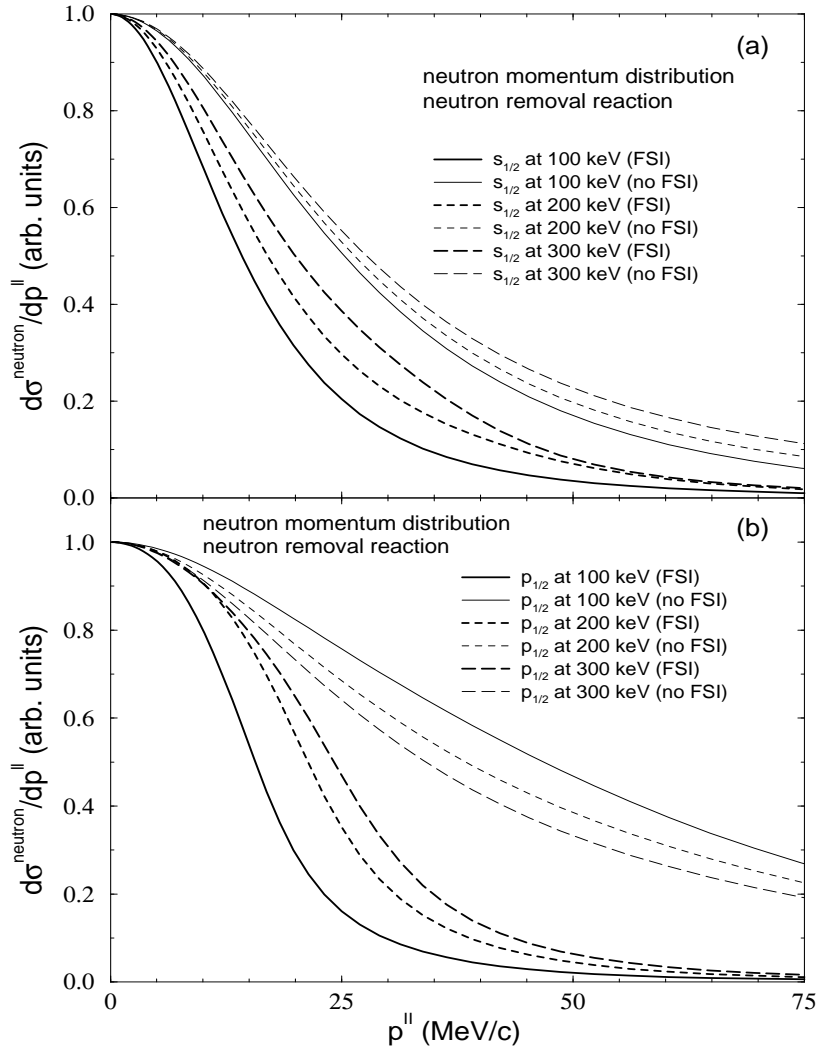}
\vspace{7.5cm}
\caption{\protect\small
The same as fig.\protect\ref{1} for one-dimensional neutron momentum
distributions.
}
\label{2}
\end{figure}

In fig.\ref{2} we show the one-dimensional neutron momentum distributions
for a neutron removal process. The notation to distinguish the curves
is the same as in fig.\ref{1}. The FWHM for these distributions are shown
in the fourth and fifth columns of table II. Let us focus first on the
momentum distributions without final state interactions (thin
curves). As in fig.\ref{1} we observe that a higher $s$-wave content in the
neutron-$^9$Li subsystem decreases the widths of the momentum
distributions. The FWHM decreases from 94 MeV/c to 51 MeV/c when the
$s$-wave content increases from 22\% to 97\%.

When final state interactions are included in the calculations (thick
curves), the change in the neutron momentum distributions is much more
significant than for core momentum distributions. For low-lying
virtual $s_{1/2}$-states (upper part of fig.\ref{2}) the width of the
distribution decreases for a virtual state at 100 keV by as much as
45\% due to final state interactions. We still observe that the lower
the virtual $s_{1/2}$-state (high $s$-wave content in the neutron-core
subsystem) the narrower the momentum distribution.  For low-lying
$p_{1/2}$-resonances in $^{10}$Li (lower part of fig.\ref{2}) we also
observe a drastic modification of the neutron momentum distributions
due to final state interactions. However, we now see that the lower
the $p_{1/2}$-resonance (higher $p$-wave content) the narrower the
momentum distribution. This result seems to contradict our experience
from the core momentum distribution where the presence of the
centrifugal barrier was decisive. But the effect of final state
interactions on the much lighter neutron in a low-lying $p$-resonance is
apparently much larger than the simultaneous broadening due to the
centrifugal barrier.

We can try to estimate the peak position of the $p$-wave contribution to
the momentum distribution. The relative momentum ($\mbox{\bf p}_r$)
between the neutron and the core in the final state preferably takes
the value $\sqrt{2 \mu E_{\mbox{\scriptsize res}}/\hbar^2}$, where
$E_{\mbox{\scriptsize res}}$ is the energy of the resonance and $\mu$
is the reduced mass of the two-body system.  The momentum of the neutron
relative to the center of mass of the projectile is very close to
$\mbox{\bf k}_x$, since the constant $b_i$ in eq.(\ref{trans}) is close
to zero due to the large mass of the core.  Then the neutron momentum
distributions in fig.\ref{2} would be concentrated around the relative
momentum $\mbox{\bf p}_r=\sqrt{\mu / m} \mbox{\bf k}_x$, see
eq.(\ref{kp}). If the resonance energy is low, the final state
interactions would try to create a narrow momentum distribution. This
argument is valid for both low-lying virtual $s_{1/2}$-states and
low-lying $p_{1/2}$-resonances. Thus, we can obtain a narrow neutron
momentum distribution with FWHM compatible with the experimental value
in table II both with low-lying $s_{1/2}$ and $p_{1/2}$-states.

For two-dimensional neutron and core momentum distributions in a
neutron removal process we obtain conclusions similar to those
extracted from figs.\ref{1} and \ref{2}. As a general rule we can say that
momentum distributions are narrower and more affected by final state
interactions when a low-lying virtual $s$-state is present in the
neutron-core subsystem (high $s$-wave content) than with a low-lying
$p$-resonance (low $s$-wave content). As for the one-dimensional
distributions the neutron momentum distributions influenced by the
final state interactions do not follow this rule. We can also get a
narrow momentum distribution with a low-lying $p_{1/2}$-resonance in
$^{10}$Li.

\begin{figure}[t]
\epsfxsize=12cm
\epsfysize=7cm
\epsfbox[600 200 1150 550]{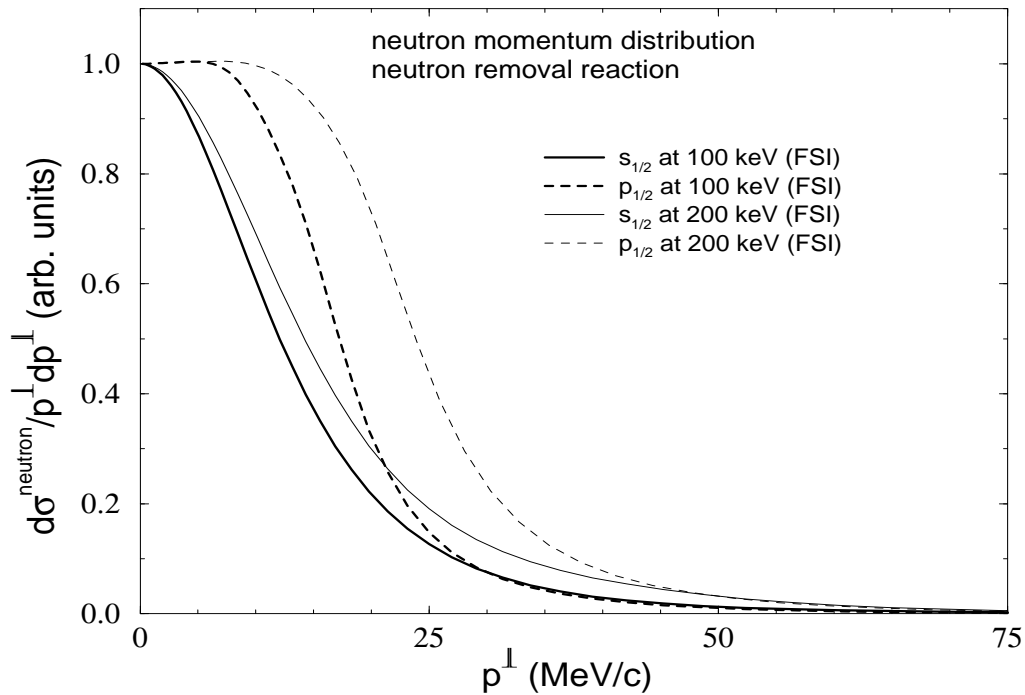}
\vspace{3.2cm}
\caption{\protect\small
Two-dimensional neutron momentum distribution for a neutron removal
process in $^{11}$Li fragmentation. FSI has been included in all the
calculations. Solid curves represent the cases of low-lying virtual
$s_{1/2}$-states at 100 keV (thick curve) and 200 keV (thin curve) in
the neutron-$^9$Li subsystem. Dashed curves represent the case of
low-lying $p_{1/2}$-resonances at 100 keV (thick curve) and 200 keV
(thin curve) in the neutron-$^9$Li subsystem. The momentum of the
neutron is referred to the center of mass of the $^{11}$Li projectile.
The $^{11}$Li spin is assumed to be zero.  } 
\label{3}
\end{figure}

In fig.\ref{3} we show the two-dimensional neutron momentum distributions for
a neutron removal process. The solid curves correspond to the cases
where a virtual $s_{1/2}$-state is present at 100 keV (thick solid
curve) and at 200 keV (thin solid curve) in the neutron-core
subsystem, while the dashed curves represent the cases where a
$p_{1/2}$-resonance is at 100 keV (thick dashed curve) and at 200 keV
(thin dashed curve). In all these cases final state interactions have
been included in the calculation.  

For a neutron-core subsystem with a $p$-resonance at a given energy
$E_{\mbox{\scriptsize res}}$ the momentum distribution (with FSI) will
exhibit a peak at the relative neutron-core momentum $\mbox{\bf
p}_r=\sqrt{\mu / m} \mbox{\bf k}_x=\sqrt{2 \mu E_{\mbox{\scriptsize
res}}/\hbar^2}$. For $E_{\mbox{\scriptsize res}}=100$ keV and 200 keV,
we then expect the peaks at $k_x \simeq 14$ MeV/c and at 19
MeV/c. In the two-dimensional neutron momentum distributions, we
integrate away the dependence on $k_x^\parallel$ and refer to the
center of mass of the projectile. Furthermore, even after including
the $s$-wave contribution, which is 22\% for the thick dashed curve
and 36\% for the thin dashed curve in fig.\ref{3}, we still can see a clear
bump in the momentum distributions corresponding to low-lying
$p$-resonances.  Thus, the effects of final state interactions on the
two-dimensional neutron momentum distributions results in shapes,
which depend very much on the structure of the low-lying state.

\begin{figure}[t]
\epsfxsize=12cm
\epsfysize=7cm
\epsfbox[-100 400 450 750]{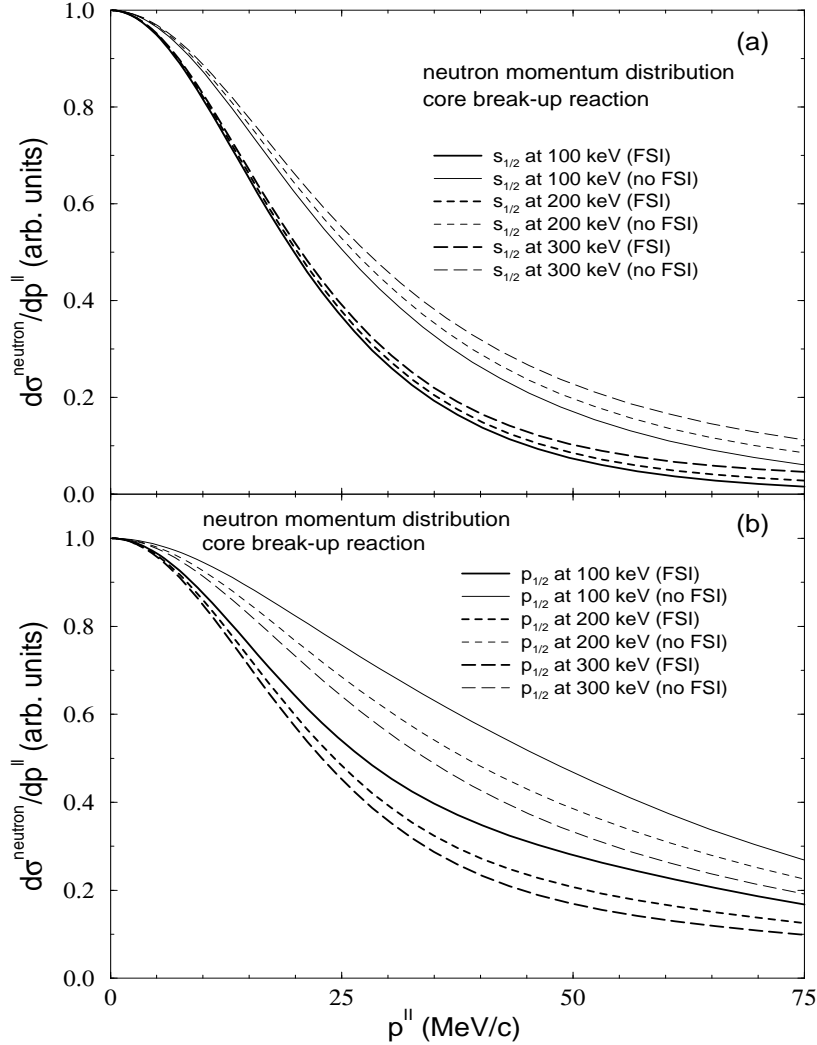}
\vspace{7.5cm}
\caption{\protect\small
The same as in fig.\protect\ref{1} for one-dimensional neutron momentum
distributions in a core break-up reaction. 
}
\label{4}
\end{figure}

Let us now turn to core break-up reactions, where the core is
violently destroyed during the interaction, but where the sudden
approximation still is valid. Then the final state interaction
(neutron-neutron interaction) is well known and the uncertainties
involved in the calculations are reduced. In fig.\ref{4} we show
one-dimensional neutron momentum distributions for core break-up
reactions. The FWHM for the curves in the figure is given in the
fourth and sixth columns of table II, where the neutron momentum
distributions without FSI are identical to those in fig.\ref{2}.

We observe that the momentum distributions are rather insensitive to
the structure of the neutron-core subsystem although the distributions
are narrower for a higher $s$-wave content. This is due to the fact
that the final state interaction (neutron-neutron interaction)
is identical in all the cases and therefore independent of the
characteristics of the neutron-core interaction.  Thus, the momentum
distributions are sensitive to the $^{10}$Li structure only through
the $^{11}$Li wave function. On the other hand, the effect produced by
the final state interaction is significant, especially for a low-lying
$p_{1/2}$-resonance in $^{10}$Li. Comparison with the experimental
width for the neutron momentum distribution in core break-up reactions
again suggests a low-lying virtual $s$-state in $^{10}$Li, see table
II.

\subsection{Momentum Distributions for Finite Core Spin}
Up to now we have neglected the spin dependence of the neutron-core
interaction or equivalently assumed that the spin of the core is zero.
Any realistic calculation should of course assume that the spin of
$^9$Li is 3/2, and the neutron-core potential should include a spin
dependence splitting the two $s$-states in $^{10}$Li into states of
total angular momenta 1 and 2. To do this, we have in the neutron-core
potential included a term proportional to $\mbox{\bf s}_n \cdot
\mbox{\bf s}_c$, where $\mbox{\bf s}_n$ is the spin of the neutron and
$\mbox{\bf s}_c$ the spin of the core. For simplicity the spin
splitting term has been introduced only in the $s$-wave, see
eq.(\ref{pots}).

The $V_s$ parameter has been chosen to be $-7.14$ MeV, the spin
splitting parameter $\gamma_s$ is taken as $-0.13$ for the solid curve
and as $0.13$ for the dashed curve. With these values the solid curves
correspond to a situation where the lowest virtual $s$-state in
$^{10}$Li has spin 1 and an energy of 50 keV, while the second
$s$-state has spin 2 and energy 340 keV. For the dashed curves the
lowest virtual $s$-state in $^{10}$Li has spin 2 and energy 97 keV,
and the $s$-state with spin 1 has an energy of 460 keV. In both cases,
the average position of the virtual $s$-states is about 230 keV. In
these calculations we have taken $V_{l=1}=-19.14$ MeV and
$V_{so}=6.97$ MeV, placing the lowest $p$-resonance in $^{10}$Li at
0.5 MeV. Final state interactions are included.

\begin{figure}[t]
\epsfxsize=12cm
\epsfysize=7cm
\epsfbox[600 200 1150 550]{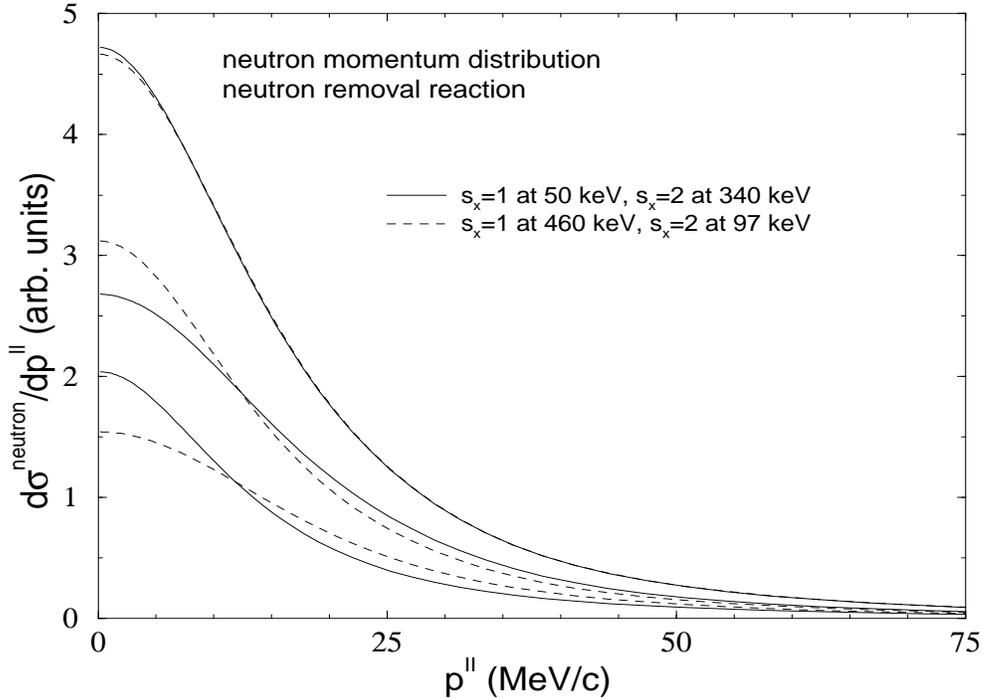}
\vspace{3.2cm}
\caption{\protect\small
One-dimensional neutron momentum distribution for a
neutron removal process in $^{11}$Li fragmentation. Solid curves: The
virtual $s$-states in $^{10}$Li are at 50 keV for $s_x=1$ and at 340
keV for $s_x=2$. Dashed curves: The virtual $s$-states in $^{10}$Li
are at 460 keV for $s_x=1$ and at 97 keV for $s_x=2$. For each case,
the lower, the middle and the upper curves are obtained assuming that the
spin of the final state system is 1, 2 and undefined, respectively.
}
\label{5}
\end{figure}

The momentum distributions can now be divided into two parts arising
respectively from angular momentum 1 and 2 of the final two-body
state. This amounts to a separation of terms with $j_x=1$ and 2 in
eq.(\ref{cm3}). In principle they can be measured independently.  In
fig.\ref{5} we plot results for the one-dimensional neutron momentum
distribution for a neutron removal process.  We have computed and
plotted three curves for each case. The lowest curve (normalized to
3/8) is the one-dimensional neutron momentum distributions assuming
that the final two-body system has spin 1.  For the next curve
(normalized to 5/8) the final two-body state is assumed to have spin
2. Finally the upper curve is the sum of the two others and therefore
normalized to 1. For a final state of spin 1, the widths of the
momentum distributions clearly increase with the energy of the virtual
$s$-state. Compare the lowest solid and dashed curves with FWHM of 26
and 37 MeV/c, respectively. This trend is also seen for a final state
of spin 2. Compare here the intermediate solid and dashed curves with
FWHM of 36 and 30 MeV/c, respectively.

We now assume that the spin of the two-body system in the final state
is not measured. Then the distributions for the individual spins should be
averaged with the statistical weighting already included in the
normalizations. This amounts to include the $s_x$ summation in
eq.(\ref{cm3}). The two cases in fig.\ref{5} resulting in the upper the
solid and dashed curves can hardly be distinguished (FWHM=32
MeV/c). This is due to the fact that the statistically weighted
average energy of the two virtual $s$-states in $^{10}$Li in both
cases is about 230 keV. The Pauli principle dictates that both
$s$-states independent of splitting are equally occupied in
$^{11}$Li. The influence of the $p$-resonance at 0.5 MeV is
insignificant and the $s$-wave content is always about 91\%. Therefore
the narrow distribution arising from one state must then be
compensated by the broader distribution from the other s-state.

Thus, when the spin of the final two-body state is unknown the
momentum distributions are only sensitive to the average position of
the virtual $s$-states and the $s$-wave content in the initial wave
function.  Of course this conclusion also holds for core momentum
distributions and core break-up reactions. If the momentum
distributions could be measured for a given spin of the final two-body
state, information about the lowest lying virtual $s$-state of $^{10}$Li
could be extracted from our analyses in a very direct way.

\section{Realistic Calculations for $^{11}$Li} \label{realistic}

The calculations we made in the previous subsection, even though a
very simple $^{11}$Li structure was assumed, indicate that the unbound
$^{10}$Li nucleus has a low-lying virtual $s$-state (table II).  A low
lying $p$-resonance clearly overestimates both the width of the core
momentum distribution and the neutron momentum distribution in a core
break-up reaction. Recent analyses suggest that a virtual $s$-state
is present at around 50 keV \cite{zinser} together with a
$p$-resonance at 0.5 MeV \cite{young2}.

We first adjust the strength parameters in the neutron-$^9$Li
potential, where $V_{l=1}$ and $V_{so}$ are chosen to place
the lowest $p$-resonance  at 0.5 MeV. The spin-orbit parameter
$V_{so}$ is used to vary the $p$-wave content in the neutron-$^9$Li
subsystem of the total $^{11}$Li wave function. This is achieved by
moving the higher-lying $p$-resonances up or down. Then the values of
the parameters $V_s$ and $\gamma_s$ are chosen to give a low-lying
virtual $s$-state and at the same time reproduce the correct binding
energy and rms radius of $^{11}$Li. The lowest virtual $s$-state in
$^{10}$Li is assumed to have spin 2, but choosing spin 1 instead does
not produce significant changes of the numerical results. The
interesting physical parameters are then energy and size of $^{11}$Li,
the $p$-wave content and the positions of the lowest $s$- and $p$-wave
resonances in $^{10}$Li.

\begin{figure}[t]
\epsfxsize=12cm
\epsfysize=7cm
\epsfbox[500 200 1050 550]{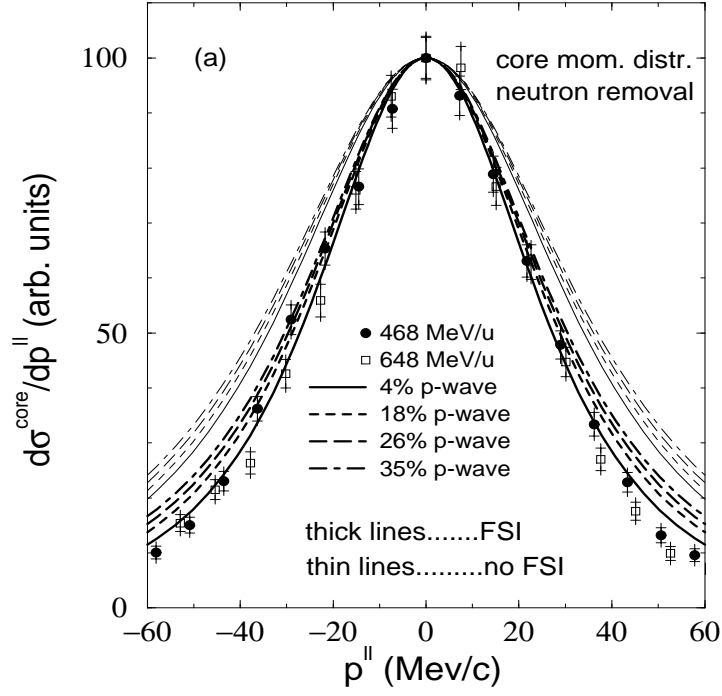}
\vspace{3.2cm}
\caption{\protect\small
One-dimensional core momentum distribution for a $^{11}$Li
fragmentation reaction. The lowest virtual $s$-state and $p$-resonance
in $^{10}$Li have an energy of 50 keV and 0.5 MeV, respectively. The
thick (thin) curves are the calculations with (without) final state
interactions.  The $p$-wave content in the neutron-$^9$Li subsystem is
4\% (solid curve), 18\% (short-dashed curve), 26\% (long-dashed curve),
and 35\% (dot-dashed curve). Experimental data are taken from
\protect\cite{geissel}.
}
\label{6a}
\end{figure}

In fig.\ref{6a} we show one-dimensional $^9$Li momentum distributions from
$^{11}$Li fragmentation. The energy of the lowest virtual $s$-state is
50 keV.  Different $p$-state contents in the neutron-$^9$Li subsystem
have been considered, 4\% (solid curves), 18\% (short-dashed curves),
26\% (long-dashed curves), and 35\% (dot-dashed curves). The thick and
the thin curves are the calculations with and without final state
interactions, respectively. The experimental data correspond to
longitudinal $^9$Li momentum distributions from a fragmentation
process of a $^{11}$Li projectile of energy 468 MeV/u and 648 MeV/u
colliding with an Al target \cite{geissel}. As expected and observed
in fig.\ref{1}, the effect of final state interactions is not very big, due
to the large mass of the core.  However, even these small effects
seems to be necessary to reproduce the measured core momentum
distributions, which in fact are rather insensitive to the structure
of the neutron-core subsystem.  All the curves including final state
interactions match the experimental data equally well.

Deviations between measured and computed curves increase with
momentum. Two different sources for this disagreement are
obvious. First the three-body model does not include core degrees of
freedom. Therefore, if the neutron-core distance is smaller than the
core-radius of about 3 fm, or equivalently the neutron-core relative
momentum is larger than about 65 MeV/c, the model is not
applicable. Secondly, only contributions from nuclear break-up
reactions are included in the sudden approximation and other processes
are assumed to be negligible. This is in general believed to be very
well fulfilled assumptions, especially for the longitudinal core
momentum distributions considered here. A convincing example is the
fragmentation of $^6$He into an $\alpha$-particle and a neutron, see
\cite{zhukov}. Our case of $^{11}$Li is expected to be more sensitive
to possible additional contributions than this very well reproduced
transverse momentum distribution of the $\alpha$-particle.  The good
agreement for the major part of the momentum range in the fig.\ref{6a} is
an aposteriori confirmation of the validity of the model assumptions,
i.e. the accuracy of the three-body wave function, the treatment of
the reaction mechanism and the incorporation of final state
interactions.

\vspace{-0.5cm}
\begin{figure}[t]
\epsfxsize=12cm
\epsfysize=7cm
\epsfbox[500 200 1050 550]{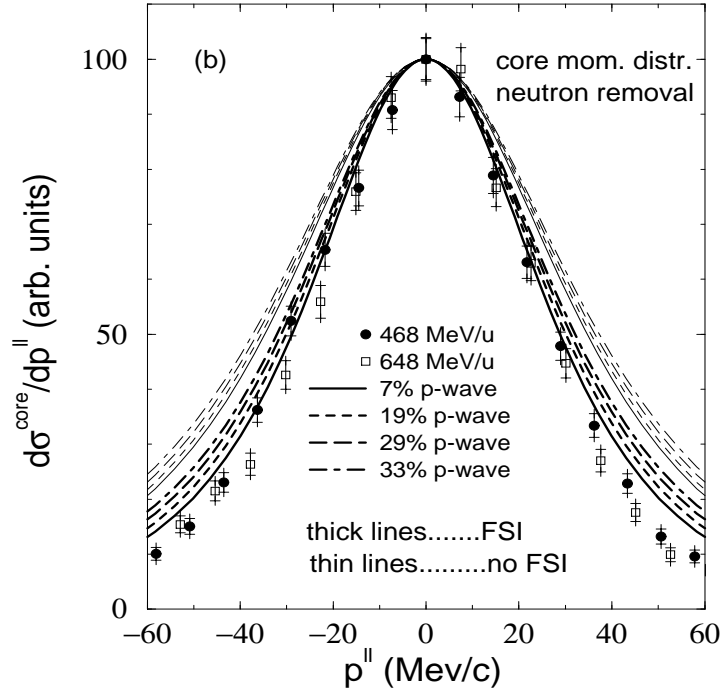}
\vspace{3.2cm}
\caption{\protect\small
Same as in fig.\protect\ref{6a} when the energy of the lowest virtual
$s$-state in $^{10}$Li is 200 keV.  The $p$-wave content in the
neutron-$^9$Li subsystem is 7\% (solid curve), 19\% (short-dashed curve),
29\% (long-dashed curve), and 33\% (dot-dashed curve).
}
\label{6b}
\end{figure}

When the energy of the lowest virtual $s$-state increases the core
momentum distributions become broader as seen in fig.\ref{6b}, where the
virtual $s$-state energy is 200 keV. The computed momentum
distributions then overestimate the width of the experimental
distributions, especially the ones with a higher $p$-wave content in
the neutron-$^9$Li subsystem. This conclusion is perhaps only
convincingly reached after both the visual impression from the figure
and consultation of the numerical widths in table II.  We can then
consider 200 keV as an upper limit to the energy of the lowest virtual
$s$-state. On the other hand, a lower limit to this energy is not
provided by the core momentum distribution, which remains unchanged
for virtual $s$-state energies below 50 keV.

\vspace{-0.5cm}
\begin{figure}[t]
\epsfxsize=12cm
\epsfysize=7cm
\epsfbox[500 200 1050 550]{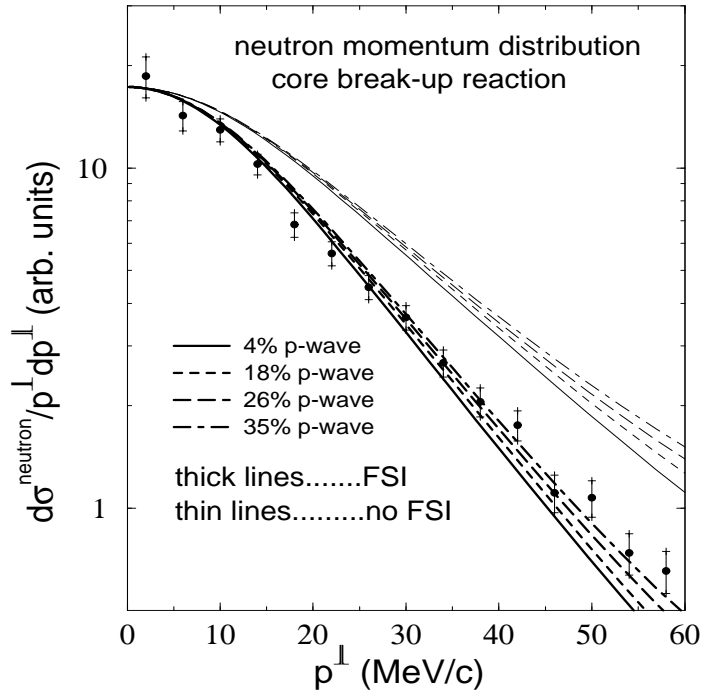}
\vspace{3.2cm}
\caption{\protect\small
Two-dimensional neutron core momentum distribution from a
core break-up reaction. The lowest virtual $s$-state and $p$-resonance
in $^{10}$Li have an energy of 50 keV and 0.5 MeV, respectively. The
interpretation of the curves is as in fig.\protect\ref{6a}. Experimental 
data are taken 
from \protect\cite{nilsson}.
}
\label{7}
\end{figure}

Let us focus now on neutron momentum distributions, where the effect
of final state interactions is larger (figs.\ref{2}, \ref{3}, 
and \ref{4}). Since for
core break-up reactions the final state nucleon-nucleon interaction is
well known, this kind of process is a good test for the model and the
method of including final state interactions. In fig.\ref{7} we plot
two-dimensional neutron momentum distributions for core break-up
reactions for the same $^{10}$Li structure as used in fig.\ref{6a} (virtual
$s$-state at 50 keV and $p$-resonance at 0.5 MeV). The meaning of the
curves is also as described in fig.\ref{6a}. The experimental data are
obtained in a fragmentation reaction with a C target and a $^{11}$Li
projectile of 280 MeV/u \cite{nilsson}. 

We observe that final state interactions now are crucial for
reproducing the experimental distribution. The calculation with
low $p$-wave content is too narrow, indicating that some $p$-state
content is needed in the neutron-core subsystem. When we gradually
increase the energy of the virtual $s$-state from 50 keV to 200 keV
the momentum distributions remain esentially unchanged (the variations
are of the size of the width of the curves). This insensitivity to the
neutron-$^{9}$Li structure is not surprising, since the remaining
particles (two neutrons) constitute a different subsystem.

\begin{figure}[t]
\epsfxsize=12cm
\epsfysize=7cm
\epsfbox[500 200 1050 550]{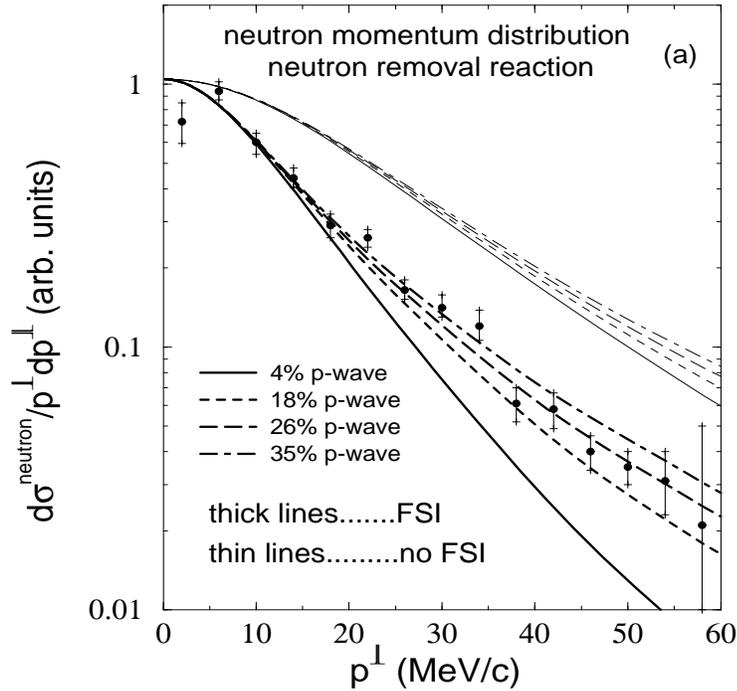}
\vspace{3.2cm}
\caption{\protect\small
Two-dimensional neutron core momentum distribution from a
neutron removal reaction. The lowest virtual $s$-state and $p$-resonance
in $^{10}$Li have an energy of 50 keV and 0.5 MeV, respectively. The
interpretation of the curves is as in fig.\protect\ref{6a}. Experimental 
data are taken 
from \protect\cite{zinser}.
}
\label{8a}
\end{figure}

Finally, we compute two-dimensional neutron momentum distributions for
a neutron removal reaction. We first consider the case of a low-lying
virtual $s$-state in $^{10}$Li with an energy of 50 keV. The
two-dimensional neutron momentum distributions are plotted in
fig.\ref{8a}. The meaning of the curves is explained in fig.\ref{6a}.  The
experimental data are taken from a fragmentation reaction with a C
target and a $^{11}$Li projectile of 280 MeV/u \cite{zinser}. We
observe again that final state interactions are decisive in order to
reproduce the experimental distribution.  Since now the final state
interaction depends on the details of the neutron-core potential, the
computed distributions show a stronger dependence on the $p$-state
content in the $^{10}$Li subsystem.  The momentum distribution with a
low $p$-wave content underestimates the width of the distribution,
while any neutron-core interaction producing more than 35\% of
$p$-wave content in $^{10}$Li overestimates the width.  The best
agreement with the experimental distribution is found for a $p$-wave
content ranging from 20 to 30\%.

\begin{figure}[t]
\epsfxsize=12cm
\epsfysize=7cm
\epsfbox[500 200 1050 550]{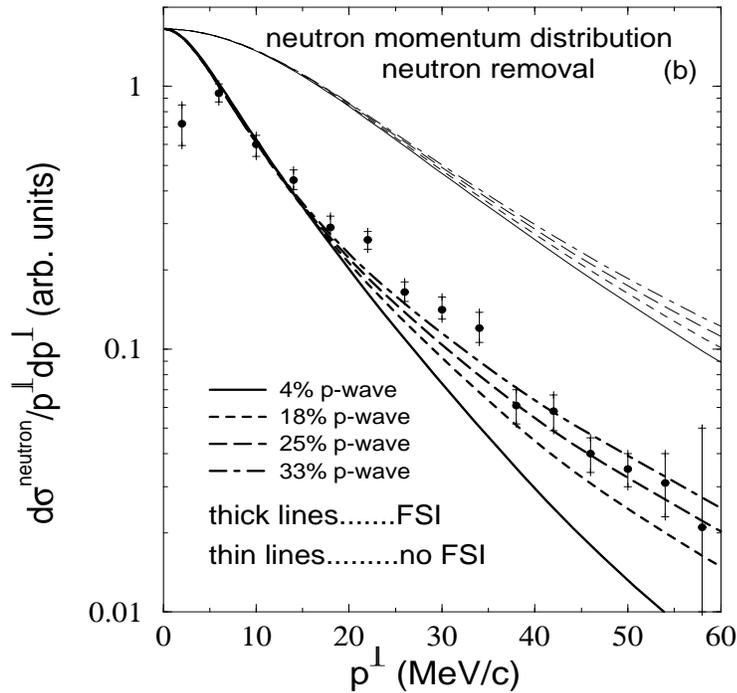}
\vspace{3.2cm}
\caption{\protect\small
Same as in fig.\protect\ref{8a} when the energy of the lowest virtual
$s$-state in $^{10}$Li is 10 keV. The $p$-wave content in the
neutron-$^9$Li subsystem is 4\% (solid curve), 18\% (short-dashed curve),
25\% (long-dashed curve), and 33\% (dot-dashed curve).
}
\label{8b}
\end{figure}

Let us now investigate how these neutron momentum distributions are
modified by a variation of the energy of the lowest virtual $s$-state
in the neutron-core subsystem. In fig.\ref{8b} we plot the same
distributions as in fig.\ref{8a} for a low-lying virtual $s$-state at 10 keV.
The momentum distributions are now much steeper at low values of the
momentum and the intermediate region is not well fitted.  None of the
distributions are able to reproduce the measured curve in the whole
range of momenta.  If we increase the $p$-wave content in the
neutron-core subsystem to fit the intermediate region, we obtain too
high values at large momenta. The best compromise would differ
considerably more both at intermediate and high momenta than the
compromise in fig.\ref{8a}. Furthermore, the deviation at very small momenta
also seems to be larger although only based on essentially one
measured point with fairly large error bars.  Thus, we conclude that
the virtual $s$-state energy in $^{10}$Li must be larger than 10 keV.

\begin{figure}[t]
\epsfxsize=12cm
\epsfysize=7cm
\epsfbox[500 200 1050 550]{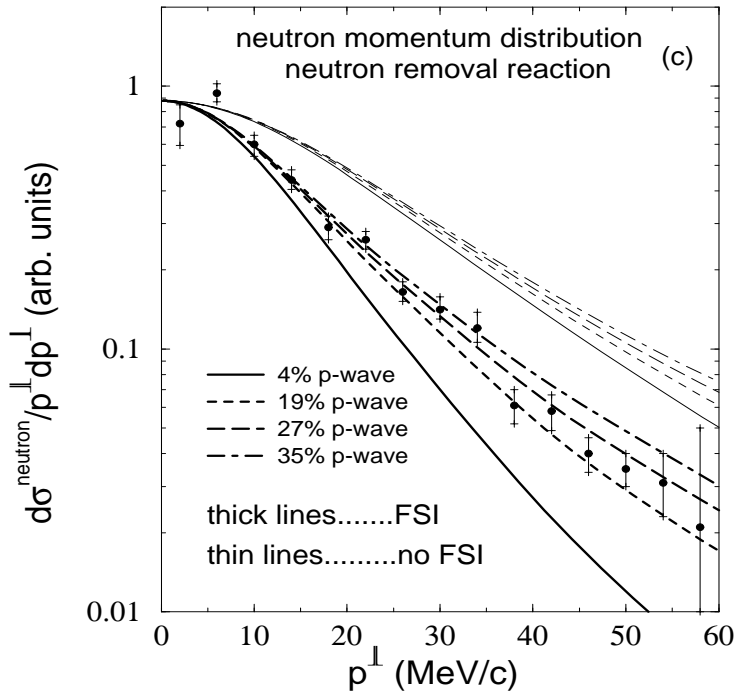}
\vspace{3.2cm}
\caption{\protect\small
Same as in fig.\protect\ref{8a} when the energy of the lowest virtual
$s$-state in $^{10}$Li is 100 keV. The $p$-wave content in the
neutron-$^9$Li subsystem is 4\% (solid curve), 19\% (short-dashed curve),
27\% (long-dashed curve), and 35\% (dot-dashed curve).
}
\label{8c}
\end{figure}

For energies of the virtual $s$-state larger than 50 keV the
experimental neutron distributions distributions are also reproduced
fairly well. In fig.\ref{8c} we show the same distributions as in 
figs.\ref{8a} and
\ref{8b}, but for a virtual $s$-state energy at 100 keV. The agreement is
improved from fig.\ref{8b} and remains almost unchanged from fig.\ref{8a}.
  Even
for higher virtual $s$-state energies the nice agreement is
maintained. This is due to the lack of accurate experimental points at
low momenta, which causes the uncertainty in the position of the
origin of the distribution.  A larger number of points in that region
would further constrain the energy of the virtual $s$-state.

Comparing figs.\ref{8a}-\ref{8c} we see that the best agreement with the
experimental neutron momentum distributions is obtained for a $p$-wave
content between 20 and 30\%. The reason is that the behaviour of the
distribution for large momenta is determined by the $p$-wave structure
of the neutron-core subsystem. This structure is essentially the same
in all these figures, i.e. the lowest $p$-resonance is placed at 0.5
MeV and the higher $p$-resonance considerably above. Furthermore, the
neutron momentum distributions are rather insensitive to the position
of the $p$-resonances. A very large variation of these energies are
needed to change the shape of the distributions significantly at
large momenta.

\section{Summary and Conclusions}
We have in this paper given a detailed formulation of a general method
to describe fragmentation of high-energy three-body Borromean halo
nuclei on light nuclear targets. In the previous papers in this series
we established general properties of the structure of three-body halo
nuclei. The most detailed information about these structures can be
obtained from measurements of momentum distributions of the fragments
from reactions of halo nuclei with various targets. The inherent
mixture of the initial halo structure and the reaction mechanism
requires a description where both ingrediences are included. The
present paper is devoted to construct a suitable method to study the
properties of fragmentation products from halo nuclei.

The initial wave function must be available with a good precision and
three-body computations are necessary. The reactions are described in
the sudden approximation, where one of the particles instantaneously
is removed from the three-body system while the other two particles
are left undisturbed. This approximation is very well adapted for the
weakly bound halo nuclei moving with high energy relative to the
targets. The remaining two particles of the projectile move roughly in
the beam direction while they continue to interact after the
collision.  This final state interaction is essential especially when
low-lying resonances are present in the two-body system. The model
must include these effects. 

We use the recently developed method to solve the Faddeev equations in
coordinate space. The initial three-body wave function is then
accurately determined. We first briefly sketch this method, which is
particularly well suited for an accurate treatment of the
large-distance behavior. The low-energy scattering properties
described by the scattering lengths are the decisive quantities
determining the large-distance behavior of the three-body
system. Therefore we parametrize the two-body interactions in terms of
simple spin-dependent operators and gaussian radial shapes. We require
that the relevant measured $s$- and $p$-wave scattering lengths as
well as the singlet $s$-wave effective range are reproduced by the
nucleon-nucleon interaction. The neutron-core interaction is adjusted
to reproduce the size and binding energy of $^{11}$Li and selected
properties of $^{10}$Li. The initial three-body wave function can now
be accurately computed.

The halo nucleus reacts with a light target nucleus and one of the
three particles is removed instantaneously. The final-state two-body
wave function is asymptotically approaching the plane wave limit, but
at smaller distances is influenced by the interaction between the
remaining particles. We include this modification by a distorted wave
approximation. This final-state two-body interaction is identically
the same as the interaction entering in the computation of the initial
three-body wave function. We maintain this consistency throughout the
paper.

The model is now completely defined and we proceed to compute the
transition matrix element which describes the probabilities for the
various break-up processes. We integrate away analytically or
numerically all non-observed quantities and end up with expressions
for the observable momentum distributions of the remaining particles.
Both one- and two-dimensional distributions are computed. To compare
with the experimental results we tranform the particle momenta to the
center of mass of the three-body halo nucleus. This is a
straightforward but technically rather lengthy procedure.

Numerical results are presented to illustrate general aspects. We
first assume a spin-less core and vary the $s$ and $p$-state
properties of the neutron-core system. The physical parameters are the
energy of the lowest virtual $s$-state, the energy of the lowest
$p$-resonance and the $p$-wave admixture in the neutron-core subsystem
of the three-body wave function. The size and binding energy of the
three-body system is left unchanged at values corresponding to
$^{11}$Li. 

Three different momentum distributions can be calculated, i.e. (i) the
core momentum distribution after neutron removal, (ii) the neutron
momentum distribution after neutron removal and (iii) the neutron
momentum distribution in a core break-up reaction. They each carry
different information. For (i) the effects of final state interactions
increase and the widths of the distributions decrease with increasing
$s$-wave content of the neutron-core subsystem. The centrifugal
barrier is crucial for this behavior.  For (ii) the final state
interaction is crucial and the momentum distributions strongly depend
on the structure of the continuum spectrum of the neutron-core
subsystem. The lower the virtual state or resonances the larger the
effect.  For (iii) the well known neutron-neutron final state
interaction has a significant effect on the distributions while the
properties of the neutron-core subsystem are less important.

The effects of finite core spin are necessary for realistic comparison
with experimental data for halo nuclei. The resulting spin splitting
of the lowest virtual $s$-states in the neutron-core system seems to
produce quantitative and perhaps even qualitative
differences. However, in most cases only the statistically weighted
average energy is important due to the constraints from the Pauli
principle, the parity conservation and the known energy of the
three-body system. One exception occurs when the two-body final-state
total angular momentum is measured in addition to the momentum
distributions. Then the angular momentum of the relative state is
determined and only one of the $s$-states contributes. The large
difference, due to final state interactions, between the momentum
distributions from the spin split states can in this way be observed.

Finally, we perform realistic calculations for $^{11}$Li and compare
with the avalaible experimental data. Since the final state
neutron-neutron interaction is well known, core break-up reactions
constitute a good test of the method used to incorporate final state
interactions in the description of the process.  The uncertainties
involved in the calculation are therefore reduced. The other type of
process, neutron removal reactions involve the structure of the
unbound system $^{10}$Li ($^9$Li +n), that simultaneously is crucial
for the structure of the $^{11}$Li three-body projectile.

The comparison with the experimental data permits us to conclude that
the final state interaction is an essential ingredient in the
computation of the momentum distributions, even for core momentum
distributions. The $p$-wave content in the neutron-$^9$Li subsystem
has been concluded to be between 20\% and 30\%. From the longitudinal
core momentum distribution and the radial neutron momentum
distribution for a neutron removal process we conclude that the energy
of the lowest virtual $s$-state in $^{10}$Li must be between 30 keV
and 200 keV. To improve the uncertainties in these conclusions, more
accurate experimental data for neutron momentum distributions for low
values of the momentum would be very helpful.

In conclusion, a consistent model has been developed to analyze
momentum distributions of the particles after high-energy
fragmentation reactions of three-body halo systems. Final state
interactions are essential. Application to $^{11}$Li and comparison
with available data led to rather severe constraints on the properties
of both $^{11}$Li and the unbound $^{10}$Li nucleus. \\

{\bf Acknowledgments} We want to thank B. Jonson and K. Riisager for
useful discussions and for making the latest experimental data
available. One of us (E.G.) acknowledges support from the European
Union through the Human Capital and Mobility program contract
nr. ERBCHBGCT930320.

\appendix
\section{Hyperspherical and hyperangular coordinates}
We consider a system of three particles with masses $m_i$ and
coordinates $\mbox{\bf r}_i$ ($i=1,2,3$). The Jacobi coordinates are
defined as
\begin{equation}
\begin{array}{ll}
\mbox{\bf x}_k=a_{ij} (\mbox{\bf r}_i-\mbox{\bf r}_j) \; , &
\mbox{\bf y}_k=a_{(ij)k} 
\left( \frac{m_i \mbox{\bf r}_i+m_j \mbox{\bf r}_j}{m_i+m_j}
-\mbox{\bf r}_k \right)  \; , \\
a_{ij}=\left( \frac{1}{m} \frac{m_i m_j}{m_i+m_j} \right)^{1/2} \; , &
a_{(ij)k}=\left( \frac{1}{m} \frac{(m_i+m_j)m_k}{m_i+m_j+m_k} \right)^{1/2}
   \; ,
\end{array}
\end{equation}
where $\{i,j,k\}$ is a cyclic permutation of $\{1,2,3\}$, and $m$ is an
arbitrary normalization mass.

From these definitions we find the momenta $\mbox{\bf k}_{x k}$ and 
$\mbox{\bf k}_{y k}$ to be:
\begin{equation}
\mbox{\bf k}_{x k}=\frac{1}{a_{ij}}  \left(
\frac{m_j}{m_i+m_j} \mbox{\bf p}_i  - \frac{m_i}{m_i+m_j} \mbox{\bf p}_j
                                   \right)  \; ,
\end{equation}
\begin{equation}
\mbox{\bf k}_{y k}=\frac{1}{a_{(ij)k}}  \left(
\frac{m_k}{m_i+m_j+m_k} (\mbox{\bf p}_i+\mbox{\bf p}_j)  - 
\frac{m_i+m_j}{m_i+m_j+m_k} \mbox{\bf p}_k
                                   \right)  \; ,
\end{equation}
where $\mbox{\bf p}_i$, $\mbox{\bf p}_j$, and $\mbox{\bf p}_k$ are the
momenta of the three particles. When these momenta are referred to the
center of mass of the three-body system ($\mbox{\bf p}_i+\mbox{\bf
p}_j+\mbox{\bf p}_k=0$) we get
\begin{eqnarray}
\mbox{\bf p}_i & = &
a_{ij} \mbox{\bf k}_{x k}   +
\frac{m_i}{m_i+m_j} a_{(ij)k} \mbox{\bf k}_{y k}  \; , \label{pi}   \\
\mbox{\bf p}_j & = &
-a_{ij} \mbox{\bf k}_{x k}   +
\frac{m_j}{m_i+m_j} a_{(ij)k} \mbox{\bf k}_{y k}  \; , \label{pj} \\
\mbox{\bf p}_k & = & -(\mbox{\bf p}_i+\mbox{\bf p}_j) 
                     = -a_{(ij)k} \mbox{\bf k}_{y k}  \; .
\end{eqnarray}

The hyperspherical variables $\{ \rho, \alpha_i, \Omega_{x i},
\Omega_{y i} \}$ are defined as
\begin{equation}
\rho=\sqrt{x_i^2+y_i^2} \; , \hspace{1cm} \alpha_i=\arctan(x_i/y_i)  \; ,
\end{equation}
where $\Omega_{x i}$ and $\Omega_{y i}$ define the directions of $\mbox{\bf
x}_i$ and $\mbox{\bf y}_i$.

Analogously we define the hyperspherical variables in momentum space
$\{\kappa, \alpha_{\kappa i}, \Omega_{k_x i}, \Omega_{k_y i} \}$ as
\begin{equation}
\kappa=\sqrt{k_{x i}^2+k_{y i}^2} \; , \hspace{1cm}
\alpha_{\kappa i} = \arctan(k_{x i}/k_{y i})  \; ,
\end{equation}
where $\Omega_{k_x i}$ and $\Omega_{k_y i}$ define the directions of 
$\mbox{\bf k}_{x i}$ and $\mbox{\bf k}_{y i}$.


\newpage

\begin{thebibliography}{99}
\bibitem{karsten} K. Riisager, A.S. Jensen, and P. M{\o}ller, Nucl. Phys.
{\bf A548}, 393 (1992).
\bibitem{dima1} D.V. Fedorov, A.S. Jensen, and K. Riisager, Phys. Lett.
{\bf B312}, 1 (1993).
\bibitem{dima2} D.V. Fedorov, A.S. Jensen, and K. Riisager, Phys. Rev.
{\bf C49}, 201 (1994).
\bibitem{dima3} D.V. Fedorov, A.S. Jensen, and K. Riisager, Phys. Rev.
{\bf C50}, 2372 (1994).
\bibitem {hansen} P.G. Hansen, A.S. Jensen, and B. Jonson, Annu. Rev.
Nucl. Part. Sci. 45, 591 (1995).
\bibitem {detraz} C. Detraz and D.J. Vieira, Annu. Rev. Nucl. Part. Sci. 39, 
407 (1989).
\bibitem{hansen2} P.G. Hansen and B. Jonson, Europhys. Lett. {\bf 4},
409 (1987).
\bibitem{johannsen} L. Johannsen, A.S. Jensen, and P.G. Hansen, Phys. Lett.
{\bf B244}, 357 (1990).
\bibitem{zhukov} M.V. Zhukov, B.V. Danilin, D.V. Fedorov, J.M. Bang,
I.J. Thompson, and J.S. Vaagen, Phys. Rep. {\bf 231}, 151 (1993).
\bibitem{kobayashi} T. Kobayashi, O. Yamakawa, K. Omata, K. Sugimoto, 
 T. Shimoda, N. Takahashi and Tanihata, Phys. Rev. Lett. {\bf60}, 2599 (1988).
\bibitem{anne} R. Anne {\it et al.}, Phys. Lett. {\bf B250}, 19 (1990).
\bibitem{orr} N.A. Orr {\it et al.}, Phys. Rev. Lett. {\bf 69}, 2050 (1992).
\bibitem{orr2} N.A. Orr {\it et al.}, Phys. Rev. {\bf C51} 3116 (1995).
\bibitem{zinser} M. Zinser {\it et al.}, Phys.~Rev. Lett. {\bf 75}, 1719
(1995).
\bibitem{nilsson} T. Nilsson {\it et al.}, Europhys. Lett. {\bf 30},
19 (1995)
\bibitem{humbert} F. Humbert {\it et al.}, Phys. Lett. {\bf B347}, 198 (1995).
\bibitem{zhukov2} M.V. Zhukov, L.V.Chulkov, D.V. Fedorov, B.V. Danilin, 
J.M. Bang, J.S. Vaagen and I.J. Thompson,   J. Phys. {\bf G20}, 201 (1994).
\bibitem{korshe} A.A. Korsheninnikov and T. Kobayashi, Nucl. Phys. {\bf
A567}, 97 (1994).
\bibitem{zhukov3} M.V. Zhukov and B. Jonson, Nucl. Phys. {\bf A589}, 1 (1995).
\bibitem{barranco} F. Barranco, E. Vigezzi, and R.A. Broglia, Phys.~Lett. 
{\bf B319}, 387 (1993).
\bibitem{dima4} D.V. Fedorov, E. Garrido, and A.S. Jensen, Phys. Rev.
{\bf C51}, 3052 (1995).
\bibitem{gar96}  E. Garrido, D.V. Fedorov and A.S. Jensen, Phys. Rev.
{\bf C}, (1996), in press.
\bibitem{bang} J.M.~Bang and I.J.~Thompson, Phys.~Lett. {\bf B279}, 201
(1992).
\bibitem{thomp} I.J.~Thompson and M.V.~Zhukov, Phys.~Rev. {\bf C49}, 1904
(1994).
\bibitem{newton} R.G. Newton, Scattering Theory of Waves and Particles,
(Springer-Verlag, N.Y., 1982), Second Edition, p.444.
\bibitem{brink} D.M. Brink and G.R. Satchler, Angular Momentum,
(Oxford University Press, London, 1962).
\bibitem{dumbrajs} O.~Dumbrajs, R.~Koch, H.~Pilkuhn, G.C.~Oades,
 H.~Behrens, J.J.~de Swart, P.~Kroll,  Nucl. Phys. {\bf B216}, 227 (1983).
\bibitem{young} B.M. Young {\it et al.}, Phys. Rev. Lett. 
{\bf 71}, 4124 (1993).
\bibitem{tanihata} I. Tanihata {\it et al.}, Phys.~Lett. {\bf B287}, 307
(1992).
\bibitem{young2} B.M. Young {\it et al.}, Phys. Rev. {\bf C49}, 279 (1994).
\bibitem{geissel} H. Geissel and W. Schwab, private communication.
\end{thebibliography}
\end{document}